\documentclass[aps,pra,reprint,twocolumn,superscriptaddress,showpacs,showkeys,superscriptaddress,longbibliography]{revtex4-2}

\usepackage{amsmath,amssymb,amstext,amsthm}
\usepackage[normalem]{ulem}
\usepackage{graphicx}
\usepackage{bm,bbold,braket}
\usepackage{natbib}
\usepackage{txfonts, comment,stmaryrd}
\usepackage{dcolumn}
\usepackage{txfonts,stmaryrd}
\usepackage[english]{babel}
\usepackage[colorlinks,bookmarks=false,citecolor=Mulberry,linkcolor=teal,urlcolor=Mulberry]{hyperref}

\usepackage{makecell}
\usepackage[usenames,dvipsnames]{xcolor}
\usepackage{array}
\usepackage{multirow}
\usepackage{adjustbox}
\newcolumntype{C}[1]{>{\centering\arraybackslash}p{#1}}

\begin{document}

\title{Disorder-averaged Qudit Dynamics}
\author{Gopal Chandra Santra}
\email{gopalchandra.santra@unitn.it}
\affiliation{Pitaevskii BEC Center and Department of Physics, University  of  Trento,  Via Sommarive 14, I-38123 Trento, Italy}
\affiliation{INFN-TIFPA, Trento Institute for Fundamental Physics and Applications,  Via Sommarive 14, I-38123 Trento, Italy}
\affiliation{Kirchhoff-Institut  f\"ur  Physik, Universit\"at  Heidelberg,  Im  Neuenheimer  Feld  227,  69120  Heidelberg,  Germany }

\author{Philipp Hauke}
\email{philipp.hauke@unitn.it}
\affiliation{Pitaevskii BEC Center and Department of Physics, University  of  Trento,  Via Sommarive 14, I-38123 Trento, Italy}
\affiliation{INFN-TIFPA, Trento Institute for Fundamental Physics and Applications,  Via Sommarive 14, I-38123 Trento, Italy}

\begin{abstract}
Understanding how physical systems are influenced by disorder is a fundamental challenge in quantum science. Addressing its effects often involves numerical averaging over a large number of samples, and it is not always easy to gain an analytical handle on exploring the effect of disorder. 
In this work, we derive exact solutions for disorder-averaged dynamics generated by any Hamiltonian that is a periodic matrix (potentially with non-trivial base, a property also called ($p,q$)-potency). Notably, this approach is independent of the initial state, exact for arbitrary evolution times, and it holds for Hermitian as well as non-Hermitian systems. 
The ensemble behavior resembles that of an open quantum system, whose decoherence function or rates are determined by the disorder distribution and the periodicity of the Hamiltonian.  
Depending on the underlying distribution, the dynamics can display non-Markovian characteristics detectable through non-Markovian witnesses. 
We illustrate the scheme for qubit and qudit systems described by (products of) spin $1/2$, spin $1$, and clock operators.  
Our methodology offers a framework to leverage disorder-averaged exact dynamics for a range of applications in quantum-information processing and beyond.

\end{abstract}

\maketitle

\section{Introduction}

The effective use of the quantum resources within an engineered quantum device requires precise control. 
However, even in cases where the device is almost perfectly isolated from its surroundings, uncontrolled and uncertain parameters can significantly restrict the accuracy and efficiency of the quantum device, a challenge shared across experimental platforms. 
The effect of the lack of certainty and control on the system's evolution is often modeled via disorder, described by random system parameters \cite{reichhardt2017disorder,bounds1987new,boschetti2022perspectives,reiner2018effects,albash2019analog,klimov2018fluctuations,brugger2022output}. 
In condensed matter, disordered systems are ubiquitous, and their fundamental and technological importance makes them a longstanding subject of research~\cite{anderson1958absence,edwards1975theory,korenblit1978ferromagnetism,mydosh1978spin,lagendijk2009fifty,smith2016many,abanin2019colloquium,vojta2019disorder}. 
In quantum-information-processing devices, the disorder may arise because of imperfect sample production,  
finite accuracy of control electronics, 
imprecise gate rotations, 
or slow drifts of system parameters.  
In this context, understanding disordered dynamics may be key for the mitigation of errors, e.g., suppressing charge-noise decoherence in superconducting charge qubits ~\cite{schreier2008suppressing}, boosting coherence via non-ergodicity~\cite{delbecq2016quantum}, determining optimal readout times, reducing gate errors~\cite{colless2018computation}, tracking drift errors in quantum processors~\cite{proctor2020detecting}, and mitigating sample-to-sample fluctuations in quantum dots~\cite{beverly2002quantum}. 
Disorder also plays a key role in quantum annealing~\cite{altshuler2010anderson,
knysh2010relevance, pino2018quantum, hauke2020perspectives}, environment enhanced quantum transport~\cite{
rebentrost2009environment,maier2019environment}, and the memory effect of disorder-induced localization can even be used to protect against errors~\cite{stark2011localization,wootton2011bringing,nandkishore2015many}.

While looking for a solution to a disordered model, often one is not interested in an individual disorder realization, either because it is too specific or because the disorder in the physical system is not sufficiently characterized. In such situations, one can derive information about the typical behavior of the quantum system by investigating disorder-averaged ensembles.
However, the dynamics of disordered systems are difficult to compute in general, and reaching a converged average often requires considerable numerical effort to acquire sufficient independent samples. In such scenarios, it can be useful to have an effective equation for describing the averaged dynamics, either in the form of a dynamical map or a differential evolution equation. Interestingly, the effective dynamics of the disorder-averaged ensemble can behave as that of an open system~\cite{gneiting2016incoherent}. 
For various cases, a master equation has been formulated that governs the disorder-averaged evolution, e.g., in the context of a qubit under spectral disorder~\cite{kropf2016effective}, 
anisotropic decoherence of qubits~\cite{chen2022effects}, 
relaxation of a qubit under coupling~\cite{lu2023sudden},
Dirac particles with mass perturbations~\cite{gneiting2020disorder},
many-body boson dynamics~\cite{kropf2020protecting},
simulations of the spin-boson model using a disordered ensemble~\cite{chen2018simulating},
transport properties in imperfect waveguides~\cite{han2019helical}, 
obtaining robust control pulses~\cite{araki2023robust},
quantum parameter estimation~\cite{ma2019improving}, %
equilibration dynamics of the Sachdev--Ye--Kitaev model~\cite{bandyopadhyay2023universal}, and operator spreading~\cite{paviglianiti2023absence}. 
For the case where the dynamical map describing the time-evolved ensemble density matrix can be inverted,  Kropf \textit{et al.}~\cite{kropf2016effective} discuss a formalism to obtain a master equation of Lindblad form using a matrix-based approach. 
However, deriving such an exact master equation is not always possible~\cite{andersson2007finding}.

Here, we discuss an analytical method to find dynamical maps to describe the exact disorder-averaged evolution dynamics. 
Our framework is valid for arbitrary times and for any unbiased disorder distribution (mean zero), without necessitating a master equation.
The Hamiltonians of the considered systems can be Hermitian or non-Hermitian, but they have to be periodic modulo a scalar factor, i.e., 
$H^p \propto \mathbb{I}$ or, more generally, $H^p \propto H^q$, with $p$ and $q$ integers such that $q<p$. 
In particular, this property covers periodic Hamiltonian matrices with non-trivial base $q$, also called $(q,p)$-potent.  
Prominent examples are tensor-product Hamiltonians of two-level systems, corresponding to involutary matrices with $p=2$ and $q=0$, or multi-level clock operators~\cite{ostlund1981incommensurate,huse1981simple}, such as  $p=3$, $q=0$. 
Periodic Hamiltonians are ubiquitous in quantum technologies, e.g., as generators of single- and multi-qubit~\cite{
sorensen1999quantum,nunnerich2024fast,nielsen2010quantum} or qudit gates in quantum computing~\cite{wang2020qudits}, as stabilizer operators in quantum error correction~\cite{gottesman1997stabilizer, lidar2013quantum, devitt2013quantum}, in error modeling of noisy quantum devices~\cite{georgopoulus2021modeling}, in noisy quantum algorithms
~\cite{shapira2003effect}, etc.

In what follows, we analytically derive the dynamical map giving the disorder-averaged density matrix for such periodic Hamiltonians at any given time and extract common structures for three example cases.   
The effective equations are characterized by several quantities relevant in quantum information theory, including various witnesses of non-Markovianity~\cite{breuer2009measure,rivas2010entanglement,breuer2016colloquium}. 
Although solving for the dynamics of each individual realization of the chosen example Hamiltonians is rather simple, they are non-self-averaging.  
Therefore, already for rather restricted evolution times, numerous independent disorder realizations can be needed to obtain converged results, in one of our example cases even up to more than $10^7$ realizations. In such scenarios, having the exact analytic results at one's disposal provides significant benefits.
In some cases, we can also derive a master equation~\cite{breuer2002theory} by inverting the underlying dynamical map,  
the result of which matches with Ref.~\cite{kropf2016effective}.
Thus, our results further our understanding of how dynamics typically associated with dissipative systems can emerge in deterministic disorder averages, and it may be useful in extracting exact behavior in simple but ubiquitous situations.

The structure of the rest of this article is as follows: Section \ref{sec:general} presents the analytical approach for deriving the disorder-averaged dynamics, supplemented by examples as well as several quantities that we use to characterize the dynamics, including
non-Markovianity witnesses. Next, we delve into three specific cases in detail: $(p=2,q=0)$, as described, e.g., by single-qubit operators or products of Pauli matrices (Sec.~\ref{sec: case I qubit}); $(p=3,q=0)$, which we exemplify through non-Hermitian qutrit operators (Sec.~\ref{sec: case II qutrit}); and $(p=3,q=1)$, describing, e.g., spin-1 operators (Sec.~\ref{sec: case III spin-1}). Finally, Sec.~\ref{sec: conclusion} concludes by highlighting the key features of the analytical method and discussing potential future research directions.

\begin{figure*}[hbt!]
    \centering
        \includegraphics[width=\textwidth,, clip, trim= 0 135 0 80]{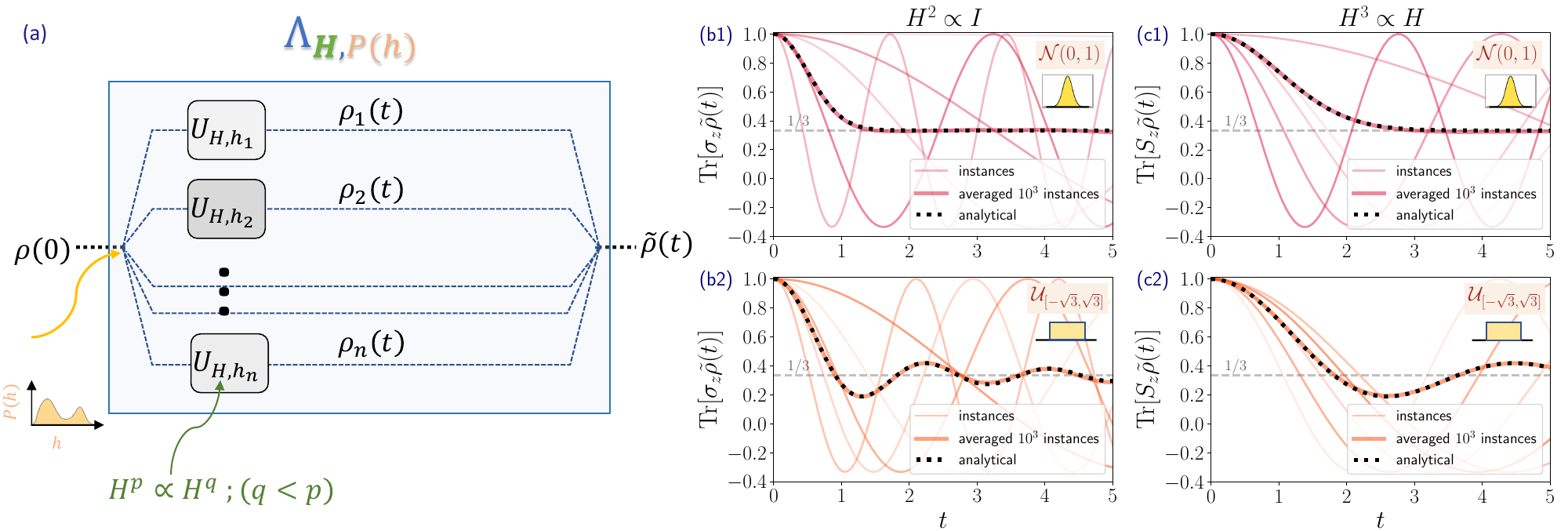}
    \caption{(a) 
    An initial state $\rho(0)$ is evolved under an ensemble of disordered Hamiltonians with a finite period ($H^p \propto H^q$, $q<p$). 
    Each instance is generated from a closed-system time evolution, with disorder $h_i$ chosen from an arbitrary disorder distribution $P(h)$.
    The disorder averaging induces the effective dynamics of $\tilde{\rho}(t)$ defined by the dynamical map $\Lambda_{H,P(h)}$,
    which is determined by the period of the Hamiltonian and the probability distribution $P(h)$ of the disorder. 
    (b1-c2) Time evolution of the expectation value of an observable [(b1,b2) $\sigma_z$, (c1,c2) $S_z$] 
    for dynamics governed by periodic Hamiltonians [(b1,b2): $H^2 \propto \mathbb{I}$, (c1,c2): $H^3 \propto H$] with Gaussian (top) and uniform (bottom) disorder distributions. The initial state corresponds to $\ket{+1}$ of the corresponding observables.
    The evolutions numerically averaged over $10^3$ samples match very well with the analytical expectation values derived with disorder-averaged state $\tilde{\rho}$.
    The averaged observables for Gaussian disorder decay monotonically (b1,c1), whereas for the uniform disorder, their expectation value oscillates, with twice the period for $H^3 \propto H$ (c2) compared to $H^2 \propto I$ (b2). Nonetheless, in all the cases, the observables reach $1/3$ for a long time.
    Directly using the averaged evolution operator can be advantageous, as it obviates the need for sufficient sampling over individual realization to reach full convergence.}
    \label{fig:intro_fig}
\end{figure*}
\section{Disorder-averaged dynamics}\label{sec:general}

To set the stage, in this section, we formally derive the dynamical map 
describing a disorder-averaged ensemble (Sec.~\ref{sec:formalsolution}). 
Finding a closed analytical solution is, in general, infeasible, but from its form, one can recognize a drastic simplification if the Hamiltonian matrix fulfills a certain structure, namely if it has a period (Sec.~\ref{sec:finiteordergeneral}). 
Before discussing several examples in detail, which we postpone to the next section, we also introduce several quantities estimating quantum informational content of the state, including some witnesses of non-Markovianity (Sec.~\ref{sec:nonmarkovianwitnesses}), which we will use in the later part of the article to characterize the disorder-averaged dynamics. 

\subsection{Disorder-averaged dynamical map: formal solution}
\label{sec:formalsolution}

Given a Hamiltonian $\hat{H}$, irrespective of it being Hermitian or non-Hermitian, and an initial state $\hat{\rho}(0)$, the final state after evolution until a time $t$ is given by 
\begin{equation}\label{eq:rhot_mat}
    \hat{\rho}(t)=e^{-i\hat{H}t} \hat{\rho}(0) e^{i \hat{H}^\dagger t}\,,
\end{equation}
(note the dagger at the right of $\hat{\rho}(0)$, necessary for the correct description of non-Hermitian dynamics~\cite{roy2023unveiling}). 
The system's dynamics follows the equation
\begin{equation}
    \partial_t \hat{\rho}(t)=-i \hat{H} \rho(t)+i \rho(t) \hat{H}^\dagger \,,
\end{equation}
which in the Hermitian case $(\hat{H}=\hat{H}^\dagger)$, turns into the well-known von Neumann equation: $\partial_t \hat{\rho}(t)=-i[\hat{H},\hat{\rho}(t) ]$.

For a general Hamiltonian $\hat{H}$, solving Eq.~\eqref{eq:rhot_mat} is not effortless. The situation becomes even more complicated if $\hat{H}=\hat{H}[h]$ describes an ensemble of Hamiltonians with a random variable $h$ over which we have to average, a procedure we denote by $\mathbb{E}[\cdot]$. In particular, one has in general $\mathbb{E}[e^{-i \hat{H}t}] \neq e^{-i\mathbb{E}[\hat{H}]t}$, making it difficult to apply the ensemble average to the evolved state. It is the aim of this paper to get a handle on the disorder-averaged density matrix for a specific but ubiquitous class of Hamiltonians useful for quantum information processing.

To calculate the disorder average irrespective of the initial state, it is advantageous to use the super-operator techniques \cite{gilchrist2009vectorization,breuer2002theory} that map between the matrix $\hat{\rho}$ and a vector $\vec{\rho}$. Any left and right multiplications to $\hat{\rho}$ can be vectorized as column vectors by the following rules 
\begin{equation} \label{eq:superoperator}
    \hat{A} \hat{\rho} \leftrightarrow (\mathbb{I} \otimes \hat{A}) \vec{\rho}    
     \quad \text{and} \quad  
     \hat{\rho} \hat{B} \leftrightarrow (B^T \otimes \mathbb{I}) \vec{\rho}\,.
\end{equation}
By series expanding on both exponentials in Eq.~\eqref{eq:rhot_mat} and using the super-operator notation above, we derive (see App.~\ref{app: first equation derivation}) 
\begin{align}\label{eq:rhot_vec}
    \vec{\rho}(t)&=\sum_{n=0}^\infty   \frac{(-it)^n}{n!}  \sum_{k=0}^{n} {n \choose k} (-\hat{H}^*)^{n-k} \otimes \hat{H}^{k} \vec{\rho}(0) \,,   
\end{align} 
where $(\ )^*$ denotes complex conjugation. 
This equation facilitates the Hamiltonian terms to be separated from the initial state by taking $\vec{\rho}(0)$ outside operator products. Such an expansion of the exponential in the power series of the density operator has previously been used to derive exact equations for the dynamics in multiple scenarios, including the evolution of a classical many-body system~\cite{weinstock1963cluster} as well as an exact generalized master equation for a quantum-mechanical system~\cite{weinstock1964generalized}, and is a procedure quite analogous to the diagrammatic cluster expansion used in statistical mechanics~\cite{kardar2007statistical}.

We denote the disorder-averaged density matrix as $\mathbb{E}[\vec{\rho}(t)]\equiv \tilde{\vec{\rho}}(t)$. Assuming no disorder in the initial state $\rho(0)$~\footnote{Note that our derivation can be extended for the disordered initial state as well, where the disorder averaging should be done together with $\rho(0)$ if it follows the same disorder distribution as the Hamiltonian or has a correlated disorder.}, we arrive at an equation where disorder averaging acts directly on different powers of the Hamiltonian and its complex conjugate,
\begin{equation}\label{eq:final eq}
    \tilde{\vec{\rho}}(t)=\sum_{n=0}^\infty   \frac{(-it)^n}{n!}  \sum_{k=0}^{n} {n \choose k} \mathbb{E}\Big[(-\hat{H}^*)^{n-k} \otimes \hat{H}^{k} \Big] \vec{\rho}(0) \equiv \Lambda_t [\vec{\rho}(0)]\,.
\end{equation}
Here, a family of linear quantum dynamical maps $\{\Lambda_t\}_{t\geq0}$ from the set of density matrices to itself formally describes the dynamics.  
In Sec.~\ref{sec: case I qubit}-\ref{sec: case III spin-1} below, we derive explicit expressions of the disorder-averaged evolution from this equation for the specific cases when the Hamiltonian is periodic, the property we introduce in the next section.  

Before that, it is illustrative to establish a formal connection of the ensemble-averaged evolution to that of an open quantum system.   
In the time-local description of dynamics in an open quantum system, the family of dynamical maps satisfies a differential equation of the form, $\partial_t\Lambda_t [\cdot]=\mathcal{L}_t \circ \Lambda_t[\cdot]$, where $\mathcal{L}_t$ is the Lindbladian that generates the dynamics~\cite{breuer2002theory}.
Alternatively, using the Nakajima--Zwanzig projective techniques, one can derive an integro-differential master equation involving a memory kernel ($\mathcal{K}$) ~\cite{breuer2002theory}. 
The general conditions on $\mathcal{L}_t$ or $\mathcal{K}$ that ensure the resulting dynamical map $\Lambda_t$ is valid, i.e., completely positive and trace-preserving, is still under research, with only certain cases fully understood (see, e.g., Refs.~\cite{budini2004stochastic,daffer2004depolarizing,maniscalco2007complete,vacchini2013nonmarkovian,vacchini2016generalized,siudzimode2017memory,witt2017exploring,viktor2019density,filippov2020phase,vacchini2020quantum,nestmann2021how}).
As a result, a general rule for constructing master equations is not always available. However, when the dynamical map is invertible, it is always possible to construct the time-local description, i.e., find $\mathcal{L}_t$~\cite{hall2014canonical}.
For example, from Eq.~\eqref{eq:final eq}, we can obtain the following equations by replacing $\vec{\rho}(0)=\Lambda_t^{-1}[\tilde{\vec{\rho}}(t)]$ in the second equality,
\begin{equation}\label{eq: lindbladian}
    \partial_t \tilde{\vec{\rho}}(t)= \partial_t \Lambda_t\Big[\vec{\rho}(0)\Big]=\partial_t\Lambda_t \circ \Lambda_t^{-1} \Big[\tilde{\vec{\rho}}(t) \Big] \equiv \mathcal{L}_t\Big[\vec{\rho}(t)\Big]\,,
\end{equation}
where $\circ$ denotes map composition, and we can identify ${\mathcal{L}_t}$ as the Lindbladian.
Kropf et al.~\cite{kropf2016effective} obtain similar master equations based on a matrix formalism~\cite{andersson2007finding,hall2014canonical} when $\Lambda_t^{-1}$ exists, which forms the foundation for the exact dynamics in their cases. Here, instead, we focus on obtaining the dynamical map $\Lambda_t$, which is enough to exactly describe the ensemble-averaged dynamics (i.e., $\tilde{\rho}(t)$), and our method is independent of the representation of the Hamiltonian in any particular basis. 
When $\Lambda_t^{-1}$ exists, we also write down the corresponding master equation. 

\subsection{Periodic Hamiltonians}
\label{sec:finiteordergeneral}
The general formal solution achieved above in Eq.~\eqref{eq:final eq} does often not help much to extract the physical properties of the system, as one has to calculate arbitrary powers of $\hat{H}$ to obtain $\tilde{\vec{\rho}}(t)$, at least when one wants to go beyond a perturbative short-time expansion. 
However, for Hamiltonian matrices with a special structure, namely, if there exists a pair of integers $p$ and $q$ (with $p>q$) for which
\begin{align}
    \hat{H}^p = h^{p-q} \hat{H}^q\,,
    \label{eq:qp-potent}
\end{align}
Eq.~\eqref{eq:final eq} can be solved exactly. 

In the literature, matrices with such a property (modulo the prefactor $h^{p-q}$) are called periodic with period $p-q$ and base $q$ \cite{li1994period} or $(q,p)$-potent \cite{jeter1982nonnegative}. 
Special cases are periodic matrices with $q=1$, called $p$-potent, and with $q=1,p=2$ (i.e., $\hat{H}^2 = \hat{H}$) called idempotent. 
If the matrix is non-singular, $\hat{H}^{p'} = h^{p'-q'} \hat{H}^{q'}$ implies the simpler relation $\hat{H}^{p} = h^{p} \mathbb{I}$, with $p=p'-q'$, which can be seen as a special case of Eq.~\eqref{eq:qp-potent} with $q=0$. 
Matrices with $q=0$ and $p=2$ are called involutary. 
Physically, the periodicity of the Hamiltonian implies the full dynamics are generated by a finite number of distinct operators.
Here, we will use this property to reduce the infinite sums in Eq.~\eqref{eq:final eq} to a finite, tractable number of terms. 

As some instructive examples, in Sec.~\ref{sec: case I qubit} to Sec.~\ref{sec: case III spin-1}, we will consider single-qubit Pauli and single-qutrit clock operators ($q=0$ and $p=2,3$) as well as spin-$1$ operators ($q=1$ and $p=3$). 
For each case, we analytically derive the dynamics of the disorder-averaged density operator and calculate several quantities to characterize it, including non-Markovianity witnesses, as described in the next section. Such analysis can be useful in determining noise channels~\cite{dutta2023qudit} and error correction protocols~\cite{ma2019improving}. 
As the spin-$1$ operator can be used as an alternative representation for qutrits~\cite{deller2023quantum}, we can also compare how the symmetry ($\mathbb{Z}_3$ for clock operators, and $\text{SU}(2)$ for spin-1) affects its dynamics. 
Before we go into the details of the results, we introduce several quantities, including witnesses of non-Markovianity, that we will use to analyze the ensuing dynamics.

\subsection{Quantum information theoretic quantities and witnesses of non-Markovianity}  
\label{sec:nonmarkovianwitnesses}
Disorder averaging can generate effective dynamics that reproduce effects typical of dissipative systems \cite{kropf2016effective,bandyopadhyay2023universal}, see Fig.~\ref{fig:intro_fig}. 
It is interesting to question whether the effective dynamics are compatible with a Markovian description. To explore this query, we will analyze several key quantities that serve as important measures in quantum information, which can also act as witnesses of non-Markovianity~\cite{rivas2014quantum, breuer2016colloquium}. 
Different and inequivalent notions of non-Markovianity exist. 
A non-Markovian process is often characterized by the backflow of information from the environment to the system, leading to revivals of information~\cite{breuer2009measure,laine2010measure}, although the precise relation between such revivals and backflow is a matter of ongoing investigations~\cite{buscemi2024information}. 
Alternatively, non-Markovianity can be described in terms of the divisibility property---positive or completely positive---of the associated dynamical maps, where non-divisibility is linked to non-Markovian behavior~\cite{hou2011alternative,wolf2008assesing,rivas2010entanglement,chruchinski2010nonmarkovian,hall2014canonical}. Non-Markovian behavior is also sometimes associated with nonexponential decays and dissipationless oscillations~\cite{zhang2012general}. Moreover, non-Markovianity can serve as a tool to probe quantum complexity~\cite{haikka2014non}. To characterize the non-Markovianity of the effective evolution, we use the following witnesses. 

\paragraph{Trace distance:}
Markovian processes tend to continuously reduce the distinguishability between any two states. Thus, according to the Breuer--Laine--Piilo (BLP) criterion, any temporal growth of distinguishability can be considered a hallmark of non-Markovianity~\cite{breuer2009measure}. As a practical witness to describe the distinguishability, we can use the trace distance between two different initial states $\rho_{1}(0)$ and $\rho_{2}(0)$, defined as  
\begin{equation}
    D(\rho_1,\rho_2;t)=\frac{1}{2} ||\rho_1-\rho_2 ||_1 \,.
\end{equation}
Here, $||M||_1$ is the trace norm $\mathrm{Tr}[\sqrt{M^\dagger M}]$. 

\paragraph{Logarithmic negativity:} 
Entanglement is considered one of the pivotal resources in quantum computing~\cite{jozsa1998quantum}. 
One of the entanglement measures, known as logarithmic negativity~\cite{plenio2005introduction,plenio2005logarithmic}, can also be used to identify non-Markovian evolutions following the Rivas--Huelga--Plenio (RHP) criterion~\cite{rivas2010entanglement}.
Consider an initial maximally entangled state between the system and an ancilla. For the example of a system given by a qubit, this state can be written as 
\begin{equation}
    \ket{\Phi}=\frac{1}{\sqrt{2}}(\ket{\uparrow\uparrow}+\ket{\downarrow\downarrow}) \Rightarrow \rho_\mathrm{SA}(0)=\ket{\Phi}\bra{\Phi}\,.
\end{equation}
Evolving the system in time with a quantum evolution operator, $\rho_\mathrm{SA}(t)=(\mathcal{E}_t \otimes \mathbb{I})\rho_\mathrm{SA}(0)$, and partially transposing the joint density matrix, one can calculate the logarithmic negativity as
\begin{equation}
E_N(\rho_\mathrm{SA})=\log_2||\rho^{T_\mathrm{S}}_\mathrm{SA}||_1 \,.
\end{equation}
Here, $(\ )^{T_\mathrm{S}}$ denotes the partial transposition with respect to the system $\mathrm{S}$.
As Markovian dynamics will always destroy quantum correlations between ancilla and system, a temporal increase in logarithmic negativity witnesses non-Markovianity~\cite{rivas2010entanglement, plenio2005logarithmic}. 

\paragraph{Purity:}

The purity of a state is defined as $\text{Tr}[\tilde{\rho}^2(t)]$. 
For finite-dimensional Hilbert spaces, the purity is monotonically decreasing during dynamics generated by a Lindbladian if and only if the Lindbladian is unital (U), i.e., $\mathcal{L}_t[\mathbb{I}]=0$
~\cite{lidar2006conditions}. Note that a Lindbladian form (where decay rates are absorbed in jump operators, thus requiring positive decay rates) guarantees a completely-positive-trace-preserving map (CPTP)~\cite{breuer2009measure,hall2014canonical}. If the dynamics induced by any unital Lindblad-like generator~\footnote{By Lindblad-like, we mean the form where the decay rate is not absorbed in the jump operators.} leads to an increasing purity, then the associated map is non-CPTP in the time-interval. Thus, the increase of the purity can witness non-Markovianity for unital maps~\cite{haseli2015measure}. The dynamics induced by the qubit and spin-1 Hamiltonians considered in Sec.~\ref{sec: case I qubit} and Sec.~\ref{sec: case III spin-1}, respectively, fall into this category, i.e., unital, while that of the non-Hermitian qutrit Hamiltonian in Sec.~\ref{sec: case II qutrit} does not (see Appendix~\ref{app: unitality}). 

We want to highlight that although purity-increasing maps can be unital, they cannot be represented in the exact Lindbladian form, as the CPTP conditions need to be conserved there. Thus, such non-Markovian features do not contradict the `if and only if' conditions for unital Lindbladians mentioned above. For example, such a scenario can appear in the presence of a negative decay rate in a Lindblad-like master equation, as discussed in detail in Sec.~\ref{sec: effective evolution}. 

To summarize, if the map is unital, an increasing purity is a good non-Markovianity witness, but if the map is non-unital, one has to employ different measures~\cite{liu2013nonunital} to reveal the non-Markovianity originating from the nonunital aspect of the dynamics. 
Regardless of the purity's ability to reveal non-Markovianity, it remains an important figure of merit, e.g., in optimal control: a control pulse can be identified as robust against imperfections when the purity of a disorder-averaged state revives near the completion of the pulse~\cite{araki2023robust,gneiting2020disorder}. The presented framework may help to directly find robust pulses from the disorder-averaged evolution, removing the need for a search from numerically averaged sample trajectories.\\

Note that all the above witnesses are non-linear functions of $\rho(t)$.
This property implies that the disorder-averaged witness, calculated from the disorder-averaged state $\tilde{\rho}$, differs from the average witness of individual disorder realizations $\rho(t)$, i.e,  $W[\tilde{\rho}] \neq \mathbb{E}[W[\rho]]$, where $W$ denotes the non-linear witness function (in contrast to standard observables that are linear functions of the state). 
Which averaging procedure is the more relevant one can depend on the situation at hand. E.g., often the disorder varies from shot to shot, making $\rho$ hardly accessible and rendering the ensemble $\tilde{\rho}$ a potentially better description of the system. 
As each independent sample describes a fully unitary evolution, $\mathbb{E}[W[\rho]]$ will return only trivial results. In contrast, as we will see further below, $W[\tilde{\rho}]$ can yield striking differences between the various disorder distributions considered. 
The non-Markovian witnesses can thus serve as a tool to illuminate the characteristics of the emergent quantum dynamics.

\section{Case I: $\hat{H}^2 = h^2 \mathbb{I}$} \label{sec: case I qubit}
In this section, we derive the exact quantum dynamics for one of the simplest possible classes covered by our framework, $\hat{H}^2 = h^2 \mathbb{I}$, i.e., involutary matrices with $p=2,q=0$. Particular examples are
\begin{enumerate}
    \item A single spin $S=1/2$ or qubit,  
    \begin{equation}\label{eq: H_1}
        \hat{H}_1=h \vec{n}\cdot \vec{\sigma}\,,
    \end{equation}
    where $\vec{n}$ is a three-dimensional unit vector and $\vec{\sigma}=(\sigma_x,\sigma_y,\sigma_z)$ is the vector of Pauli matrices. For concreteness, in what follows, we choose without the restriction of generality $\vec{n}=(1,1,1)/\sqrt{3}$;
    
    \item Many-body tensor products of Pauli matrices 
    \begin{equation}
        \hat{H}_N= h \bigotimes_{i=1}^N (\hat{H}_{1i}/h)\,; \label{eq: Hamiltonian_Hn}
    \end{equation}

    \item Tensor products of $\sigma_{\alpha}$ at different positions, $P_{\alpha}$, where $\sigma_{\alpha}=\{\mathbb{I}, \sigma_x, \sigma_y, \sigma_z \}$. For example, $\sigma_x \otimes \mathbb{I} \otimes \sigma_y$.
\end{enumerate}

\indent Such Hamiltonians are ubiquitous in quantum mechanics, ranging from textbook examples such as Stern--Gerlach experiments on single spins to building blocks of advanced quantum technologies in cold-atom and quantum-computing platforms (e.g., single qubit operations are described by $H_1$, while two-qubit gates such as the Mølmer–-Sørensen gate are described by $H_2$ \cite{wang2001multibit}). The parity checking operator for quantum error detection and corrections belongs to $P_{\alpha}$~\cite{lidar2013quantum}.
Here, we want to analyze analytically what effects disorder has on such systems in general.

\subsection{Effective evolution and master's equations}\label{sec: effective evolution}

We only consider mean-zero disorder distributions. As a consequence, $\mathbb{E}[h^{2m+1}]=0$ with $m\in \mathbb{N}$, and we are left with the terms of even $n$'s of Eq.~\eqref{eq:final eq}. On top of that, as $\hat{H}^2 = h^2 \mathbb{I}$, all powers of $\hat{H}$ are either proportional to $\mathbb{I}$ (even) or $\hat{H}$ (odd). These constitute the dynamics of $\tilde{\rho}(t)=\mathbb{E}[\rho(t)]$ as (see App.~\ref{app:derivation_qubit} for details)
\begin{equation}
    \label{eq:rho_qubit}
    \tilde{\vec{\rho}}(t)=\Bigg[ \left(\frac{1+G(t)}{2}\right) \mathbb{I} \otimes \mathbb{I} + \left( \frac{1-G(t)}{2}\right)\tilde{H}^* \otimes  \tilde{H}   \Bigg]\ \vec{\rho}(0) \equiv \Lambda_t \vec{\rho}(0)\,,
\end{equation}
where $\tilde{H}=\hat{H}/h$ is the dimensionless Hamiltonian. 
Note that Eq.~\eqref{eq:rho_qubit} needs only two operators out of four possible combinations possible from $\{\mathbb{I}$ and $\tilde{H}\}$ because both $\mathbb{I} \otimes \tilde{H}$ and $ \tilde{H}^* \otimes \mathbb{I}$ are associated to odd powers of $h$, and thus vanish after disorder averaging for a distribution with mean $0$. 

The general form in Eq.~\eqref{eq:rho_qubit} is independent of the type of disorder distribution, whose entire effect is encapsulated in the time-dependent function $G(t)$. 
For the present case of $p=2, q=0$, we have $G(t)=\mathbb{E}[e^{-2iht}]= \int dh p(h) e^{-2iht}$, i.e., it is given by the Fourier transform of the probability distribution with time $t'=2t$, and thus corresponds to the characteristic function containing the behavior and properties of the probability distribution. 
In Tab.~\ref{tab:G(t)_qubit}, we give the exact form of $G(t)$ for Gaussian and homogeneous uniform disorder distributions. 

To illustrate the accuracy of our analytical method, we consider the Hamiltonian $H_1$ given above, with $\tilde{H}=\vec{n}\cdot \vec{\sigma}$, and calculate the average magnetization ${\rm Tr}[\sigma_z \tilde{\rho}(t)]$ using the disorder-averaged state $\tilde{\rho}(t)$ from the analytical solution. 
For the considered initial state $\ket{\uparrow}=(1,0)^T$, the analytical calculation gives
\begin{equation}\label{eq: qubit_magnetization}
    {\rm Tr}[\sigma_z \tilde{\rho}(t)]=\frac{1}{3}\left(1+2G(t)\right)\,.
\end{equation}
In Fig.~\ref{fig:intro_fig}(b), we plot the cases of Gaussian disorder (top) as well as uniform disorder (bottom). 
For a reasonable comparison, we set both the variances of the uniform and the Gaussian distribution to unity, i.e., $b^2/3=1$ and $\sigma^2=1$. 
While the ensemble-averaged magnetization dynamics for the Gaussian disorder decays monotonically versus the value corresponding to an infinite-temperature state, the dynamics for other distributions can be more complicated, illustrated by the oscillations visible in the case of uniform disorder.   
For both types of distributions, the exact dynamics over the considered time range matches very well with the numerically calculated disorder-averaged magnetization of $10^3$ instances.

Furthermore, from Eq.~\eqref{eq:rho_qubit}, we can derive a dynamical master equation for the disorder-averaged density matrix following Eq.~\eqref{eq: lindbladian}. To this end, by inverting the dynamical map in Eq.~\eqref{eq:rho_qubit} (see App.~\ref{app:derivation_qubit}), we obtain a time-local master equation 
\begin{equation}\label{eq: qubit master equation}
    \partial_t \tilde{\rho}(t)=-\frac{\partial_t G(t)}{2 G(t)}\Bigg[\tilde{H} \ \tilde{\rho}(t) \ \tilde{H}^\dagger-\frac{1}{2} \bigg\{(\tilde{H})^\dagger  \tilde{H}\ ,\ \tilde{\rho}(t) \bigg\}    \Bigg]\,.
\end{equation}
From this equation, it becomes apparent that the $\tilde{H}$ coupled to the disordered field assumes the role of jump operators of a corresponding master equation in Lindblad form. Moreover, the effective decay rate $\gamma(t)\equiv -\frac{\partial_t G(t)}{2 G(t)}$ is given entirely by the temporal rate of change of the time-dependent function $G(t)$, which in turn is determined by the disorder distribution (see Tab.~\ref{tab:G(t)_qubit} for explicit forms).

For Gaussian disorder, $G(t)$ and $\gamma(t)$ are always non-negative analytic functions (see Tab~\ref{tab:G(t)_qubit}). In contrast, for uniform disorder, $G(t)$ periodically switches sign, and $\gamma(t)$ can have negative values [see Fig.~\ref{fig:plot H1}(b)]. Such dynamics can no longer be $P-$divisible~\cite{kossakowski1972quantum}, and hence it is also not $CP-$ divisible (see more in Ref.~\cite{breuer2016colloquium}), and thus a negative decay rate can also witness non-Markovianity~\cite{hall2014canonical}, as we discuss in more detail in Sec.~\ref{sec: non-markovianity_qubits}.

\begin{table}[]
     \renewcommand{\arraystretch}{1.5}
     \centering
     \begin{tabular}{|c| c| c | } 
      \hline 
       Quantities  & Gaussian $\mathcal{N}(0,\sigma^2)$ & Uniform  $\mathcal{U}_{[-b,b]}$\\
       \hline  \hline 
       Prob.\ dist.\ function: $P(h)$ & $\frac{1}{\sigma \sqrt{2\pi}} e^{-\frac{1}{2}h^2 / \sigma^2 }$ & $\frac{1}{2b}$\\ 
       Characteristic function: $\varphi_x (t')$ & $e^{-\sigma^2 t'^2 /2}$ & $\frac{\sin{bt'}}{bt'}$      \\ 
       Disorder moments: $\mathbb{E}[h^{2n}]$ & $\sigma^{2n}(2n-1)!!$ & $b^{2n}/(2n+1)$\\ 
       
       Time-dependent function: $G(t)$ &$ e^{-2\sigma^2 t^2}$ & $\frac{\sin{2bt}}{2bt}$\\
       
       Decay rate: $\gamma(t)$ &$2 \sigma^2 t$ & $\frac{1}{2}\Big(\frac{1}{t}-2b \cot{2bt}\Big)$  \\
       \hline 
     \end{tabular}
     \caption{Relevant quantities describing the exact evolution depending on the disorder distributions: probability distribution function, the characteristic function of the distribution, disorder-averaged moments, $G(t)$, emergent decay rate governing the master equation \eqref{eq: qubit master equation}. The last two quantities are specific for the case $p=2, q=0$, such as the single-qubit Hamiltonian $H_1$. }
     \label{tab:G(t)_qubit}
\end{table}
\subsection{Relation to error channels}

Quantum states are extremely fragile and can easily be disrupted by environmental interactions~\cite{zurek2003decoherence,
lopez2019all}, hardware imperfections, or operational gate errors~\cite{gardas2018defects}. Adequately understanding the dominant error channels of a given quantum hardware is essential for designing fault-tolerant architectures and error-correction codes, making it one of the critical steps in building reliable quantum computing firmware.
To estimate the effects of such sources on quantum devices, one typically constructs noise models using quantum channels, e.g., depolarizing, dephasing, amplitude damping, etc. 
The present framework can be used to derive analytic descriptions of error channels that originate from uncontrolled gate operations. 

To illustrate the main idea, consider a generic dephasing channel defined by 
\begin{equation}\label{eq: dephasing}
    \rho \rightarrow \rho'= (1-p_d) \rho + p_d\sigma_z\rho \sigma_z\,,
\end{equation}
where with probability $p_d$ the state dephases, while with probability $(1-p_d)$ it remains unchanged. Interestingly, Eq.~\eqref{eq: dephasing} coincides with
Eq.~\eqref{eq:rho_qubit} when setting $\tilde{H}=\sigma_z$ and choosing the dephasing probability $p_d=\frac{1-G(t)}{2}$. As mentioned above, $G(t)$ is the ensemble-averaged rotation generated by the disorder, $G(t)= \mathbb{E}[e^{-2iht}]$. Thus, if the system is subject to uncontrolled disorder, one can immediately determine the associated noise channel and its probability $p_d$ by only characterizing the type of disorder, here $\mathrm{exp}(-i ht \sigma_z)$,  and the distribution of the pulse areas $ht$. 
This knowledge of disorder can further be used to determine 
suitable parameters for counteracting disorder-induced dephasing~\cite{kropf2020protecting}. 
Similar considerations also hold for qudit systems~\cite{andersson2007finding, dutta2023qudit}.

\subsection{Non-Markovianity from disorder distributions}\label{sec: non-markovianity_qubits}
The analytical form of the disorder-averaged state [Eq.~\eqref{eq:rho_qubit}] permits us to analytically compute the witnesses of non-Markovianity introduced in Sec.~\ref{sec:nonmarkovianwitnesses}.  
As detailed in Tab.~\ref{tab: all_observables}, they can be expressed exactly as a function of $G(t)$, starting from any initial state, for example, --

\begin{equation} \label{eq: non-markovianity qubit}
    \begin{aligned}
        {\rm Tr}(\tilde{\rho}^2) & =\frac{1}{3}(2+G^2(t))\,, \\
        D(\tilde{\rho}_1, \tilde{\rho}_2;t) & =\sqrt{\frac{1+2G^2(t)}{3}}\,,\\
        E_N(\tilde{\rho}_{\rm SA}) & =\log_2(1+|G(t)|)\,.
    \end{aligned}
\end{equation}
In order to see the difference between Gaussian and uniform distribution, we insert the corresponding form of $G(t)$ as detailed in Eq.~\eqref{eq: non-markovianity qubit} into the witness functions. 
 
\begin{figure}[hbt!]
    \centering
    \includegraphics[width=\columnwidth]{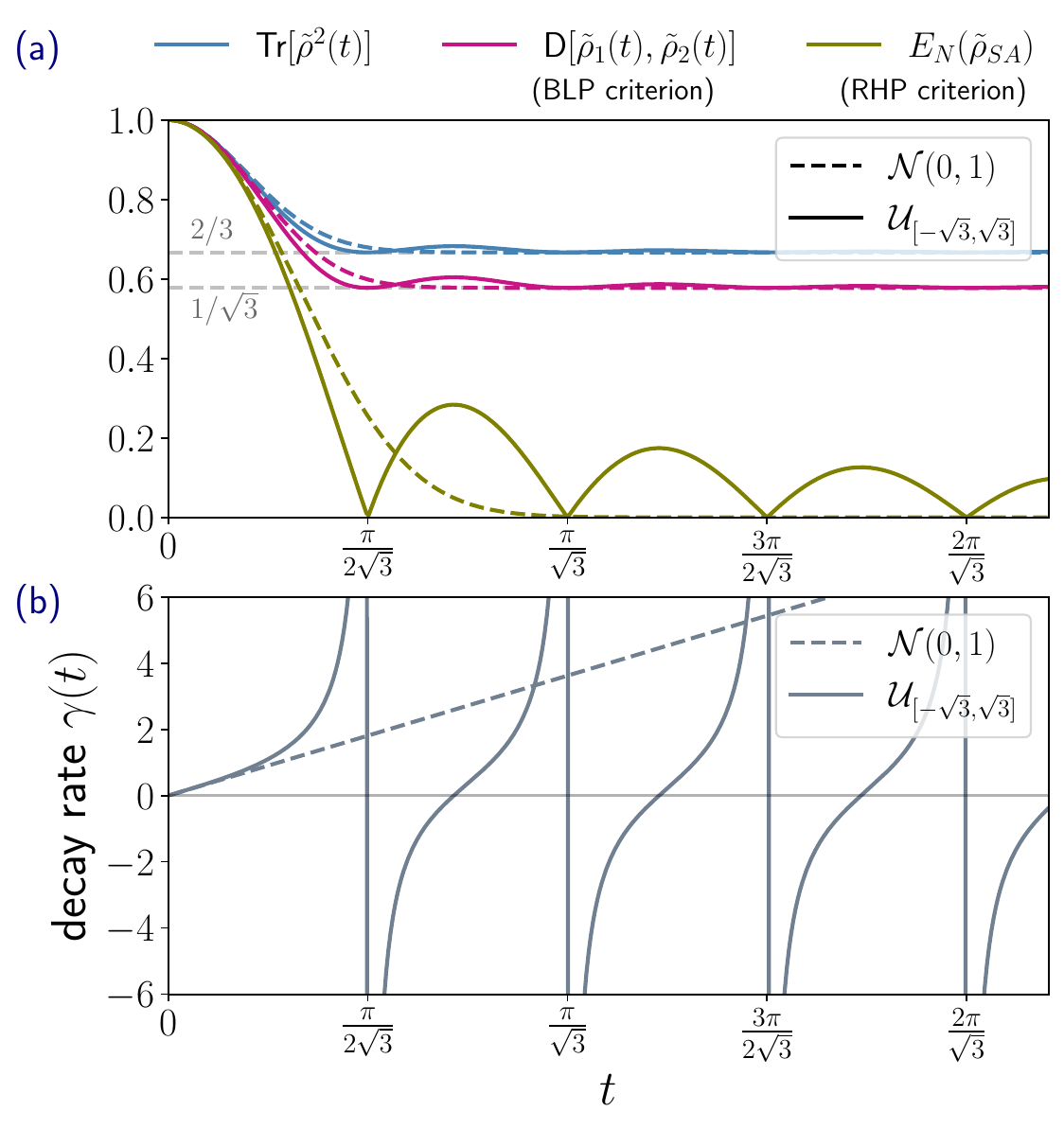}
    \caption{ 
    (a) Non-Markovian witnesses evaluated for the time-evolved disorder-averaged state for the case $p=2,q=0$ (qubit Hamiltonian) for Gaussian (dashed line) and uniform (solid line) distribution: Purity (blue), trace distance (violet), and entanglement negativity (green). All quantities are consistent with Markovian behavior for Gaussian disorder, but they show distinct non-Markovian features (revivals) for uniform disorder with a period of $\pi/2b$. (b) The decay rate in the Master Eq.~\eqref{eq: qubit master equation} also detects non-Markovianity. For the Gaussian disorder, it is always positive, indicating Markovianity, while for the uniform disorder, it assumes negative values in regions where the non-Markovian witnesses show revivals.}
    \label{fig:plot H1}
\end{figure}

With Gaussian disorder, purity, trace distance, and entanglement negativity decrease and saturate to a particular constant (see Fig.~\ref{fig:plot H1}). This behavior is compatible with a complete Markovianity of the evolution. In contrast, with uniform disorder, the non-Markovian witnesses show significant oscillations and only at late times settle to the same long-time value as that obtained for Gaussian disorder. 
The periodicity of the oscillation is determined by the periodicity of $|G(t)|$, i.e., $\frac{\pi}{2b}$. 
One can define the strength of non-Markovianity as the number of times the revival of these witnesses happens. This number is inversely proportional to the periodicity of $|G(t)|$ and thus proportional to the standard deviation of the disorder distribution~\cite{lorenzo2017quantum}. 
Thus, the revival frequency can be controlled by controlling the standard deviation of the disorder, which might be exploited for information-processing tasks~\cite{xu2013experimental,reich2015exploiting}. Moreover, with non-Markovian features, the evolution dynamics can be sped up~\cite{deffner2013quantum}, or the capacity of a quantum channel can be enhanced~\cite{bylicka2014non}. Note that, in this case, since the dynamical map is unital (see Appendix~\ref{app: unitality}), purity can be used as a witness of non-Markovianity. It functions just as effectively as other witnesses and exhibits similar behavior.

\subsection{Multi-qubit operators}

\begin{figure}[hbt!]
    \centering

    \includegraphics[width=0.9\columnwidth, clip, trim=15 50 530 55]{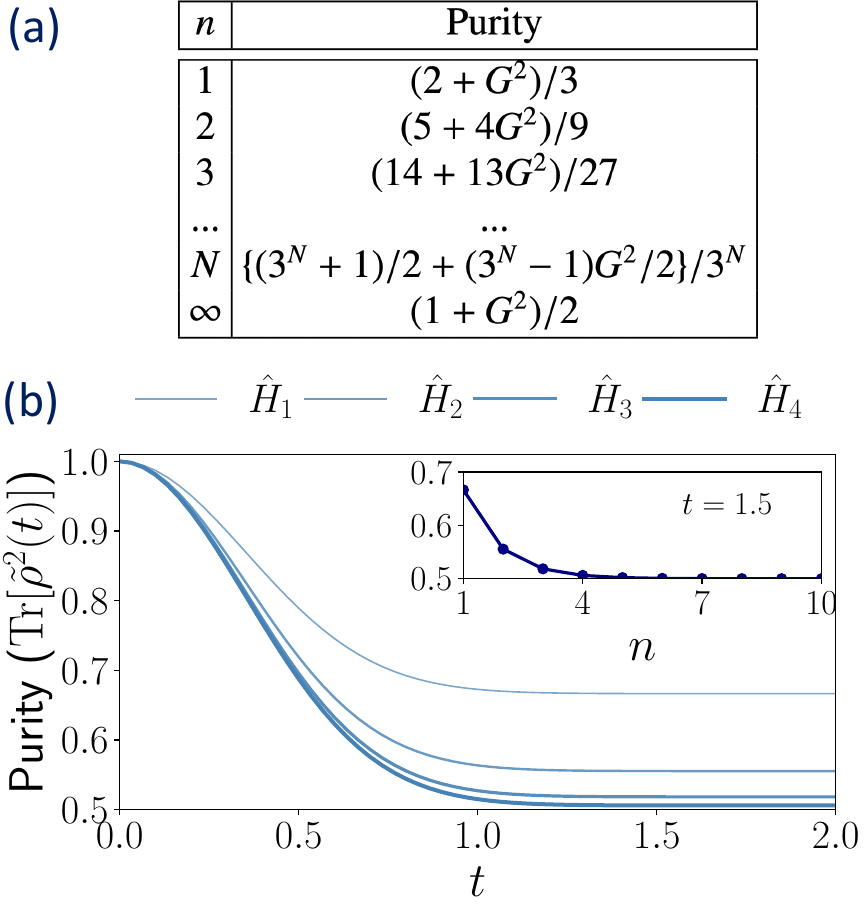}
    \caption{(a) Exact formulas for the purity of the disorder-averaged state under the evolution of $\hat{H}_N$ for initial state $\ket{\uparrow}^{\otimes N}$. The ratio between the terms proportional to $\mathbb{I} \otimes \mathbb{I}$ and $\hat{H}^T \otimes \hat{H}$ is $(3^N+1)/(3^N-1)$, which approaches $1$ as $N$ increases. (b) Exact dynamics of the purity for $N=1,2,3,4$ with Gaussian disorder. As $N$ increases, the final purity plateau reaches closer to $1/2$. } 
    \label{fig:manyqubits}
\end{figure}

A multi-qubit operator such as $\hat{H}_N$ as given in Eq.~\eqref{eq: Hamiltonian_Hn} 
also fulfills $\hat{H}^2 \propto \mathbb{I}$. Thus, the same analytical solutions as above hold. For concreteness, we consider the multi-qubit Hamiltonian
$\hat{H}_N=h \bigotimes_{i=1}^N \Big(\frac{1}{\sqrt{3}}\sum_{\alpha=x,y,z} \sigma_\alpha \Big)_i$. To illustrate the effect of increasing $N$, we calculate the purity for Gaussian disorder.  

As the table in Fig.~\ref{fig:manyqubits}(a) details, the purity consists of a time-independent constant and a time-dependent part given by $G(t)$. With increasing $N$, the ratio between the corresponding coefficients decreases, e.g., from $2$ for $N=2$ to $5/4$ for $N=3$. The general ratio is $(3^N+1)/(3^N-1)$, showing that both contributions attain equal weight at $N\to\infty$.  
As $N$ increases, the purity saturates at a lower value (see Fig.~\ref{fig:manyqubits}) and ultimately reaches $1/2$ (see inset of Fig.~\ref{fig:manyqubits}). (Due to the special form of $\hat{H}_N$, the dynamics is non-ergodic, and the saturation value remains far from $1/2^N$, the value corresponding to a maximally mixed state.)

\section{Case II: $\hat{H}^3 =h^3 \mathbb{I}$}\label{sec: case II qutrit}

Until now, we have focused on Hamiltonians generating gates on single or multiple qubits. 
However, there is currently a significant drive towards developing the use of quantum-information carriers with more than $d>2$ levels, so-called qudits~\cite{kasper2021universal,blok2021quantum,ringbauer2022universal, chi2022programmable, hrmo2023native, fischer2023universal,
calajo2024digital,gonzalez2022hardware}. 
The basic operations for qudit computations~\cite{wang2020qudits} are typically defined via the generalized Pauli operators, also known as Potts or clock operators in a statistical physics context. 
These operators present examples of $(q,p)$-potent Hamiltonians with $p=d, q=0$. 

Qudit operations can be defined by two generators, the shift operator $\tau$, which moves the qudit state from $\ket{i}$ to $\ket{i-1}$, and the phase operator $\sigma$, which feeds $d$-different eigenvalues that are defined by the $d-$th root of unity, $\omega \equiv e^{2\pi i/n}$. 
In matrix form, they read 
\begin{equation}
    \sigma=\begin{pmatrix}
      1 & 0 & 0 & ... & 0 \\
      0 & \omega & 0 & ... & 0 \\
      0 & 0 & \omega^2 & ... & 0 \\
      \vdots & \vdots & & \vdots \\
      0 & 0 & 0 & ... & \omega^{n-1} 
      \end{pmatrix}, \quad 
      \tau= \begin{pmatrix}
      0 & 0  & ...&0 & 1 \\
      1 & 0  & ...&0 & 0 \\
      0 & 1  & ...&0 & 0 \\
      \vdots & \vdots & &0 & \vdots \\
      0 & 0 &  ... &1 & 0    
      \end{pmatrix}\,, 
\end{equation}
and they fullfil the relations 
\begin{align}
    \sigma^n=\tau^n=1, \ \sigma^\dagger=\sigma^{n-1}, \ \tau^\dagger=\tau^{n-1},\ \sigma\tau=\omega \tau\sigma\,. 
\end{align}
Notice that the operators $\sigma, \tau$ are both non-Hermitian. Such non-Hermitian operators have been considered, e.g., in constructing a model of Feynman’s clock~\cite{tempel2014feynman} and in the context of topological insulators~\cite{ezawa2022clock}. Another example of such operators appears also in the context of $\mathbb{Z}_d$ parafermions, which have instead non-local commutation relations and can also be used for quantum computing~\cite{hutter2016quantum}. Recently, effective non-Hermitian qudit Hamiltonians have been realized in trapped ions~\cite{chen2024quantum,li2024programmable}.

In this paper, for the analytical calculation's tractability, we stick to the case of qutrits, i.e., $d=3$.  
We consider a generic single qutrit Hamiltonian as a sum of both shift and phase operator, 
\begin{equation}
    \hat{H}=\frac{h}{2^{1/3}}(\sigma+\tau)=h \tilde{H}\,.
\end{equation}
This Hamiltonian fullfils $\hat{H}^3 = h^3 \mathbb{I}$. 
In this case, due to the existence of three independent operators $\{\mathbb{I}, \tilde{H}, \tilde{H}^2 \}$, we can have at most $9$ possible operators contributing to the evolution of the state $\tilde{\rho}(t)$. Unlike the $H^2 \propto \mathbb{I}$ case, where the terms of odd order in $h$ vanish, in the present case all the $9$ terms contribute 
(see App.~\ref{app:derivation_qutrit} for more details). 
For simplicity, we focus on Gaussian disorder, for which we can write the final disorder-averaged state $\tilde{\rho}(t)$ as 

\begin{equation}
    \label{eq:qutrit_rhot}
    \begin{aligned}
        \tilde{\vec{\rho}}(t) 
        & =\frac{1}{9} \Big[ 3\left(\mathbb{I} \otimes \mathbb{I}+\tilde{H}^{*2}\otimes \tilde{H} +\tilde{H}^{*}\otimes \tilde{H}^2 \right) \\
        &+ \underbrace{\vec{v}(t) \cdot (1,1,1)^T}_{G_1(t)} \left( 2\mathbb{I} \otimes \mathbb{I} -  \tilde{H}^{*2}\otimes \tilde{H}-\tilde{H}^* \otimes \tilde{H}^{2} \right)\\
        &+ \underbrace{\vec{v}(t) \cdot (1,\omega^2,\omega)^T}_{G_2(t)} \left(2{\tilde{H}^{*}\otimes \tilde{H}} -  \tilde{H}^{*2}\otimes \mathbb{I}-\mathbb{I} \otimes \tilde{H}^{2}\right) \\ 
        &+ \underbrace{\vec{v}(t) \cdot (1,\omega,\omega^2)^T}_{G_3(t)} \left(2\tilde{H}^{*2}\otimes \tilde{H}^2 -  \tilde{H}^{*}\otimes \mathbb{I}   -\mathbb{I} \otimes \tilde{H} \right) \Big]\  \vec{\rho}(0)\,,
    \end{aligned}
\end{equation}
where for convenience we have defined the vector $\vec{v}(t) \equiv (e^{\frac{3}{2}\sigma^2 t^2},e^{\frac{3}{2} \omega \sigma^2 t^2},e^{\frac{3}{2}\omega ^2 \sigma^2 t^2})$. 
Note, for a non-Hermitian system, the state $\tilde{\rho}$ has to be normalized by ${\rm Tr}(\tilde{\rho})$ in order to obtain normalized probabilities.

To see how this analytical equation regulates the dynamics, we first look at $t=0$, where $\vec{v}(0)=(1,1,1)$. In this limit, only the first two lines of Eq.~\eqref{eq:qutrit_rhot} survive and add up to give $\mathbb{I}\otimes \mathbb{I}\ \vec{\rho}(0)$, as expected. 
Similar to the qubit case, we can also identify the time-dependent functions $G_1(t), G_2(t), G_3(t)$ (see Tab.~\ref{tab: all_propagator}), which govern the full dynamics at $t>0$ and which differ only by the combinations of coefficients $\{1,\omega,\omega^2\}$. 

\begin{figure}[hbt!]
    \centering
     \includegraphics[width=\columnwidth]{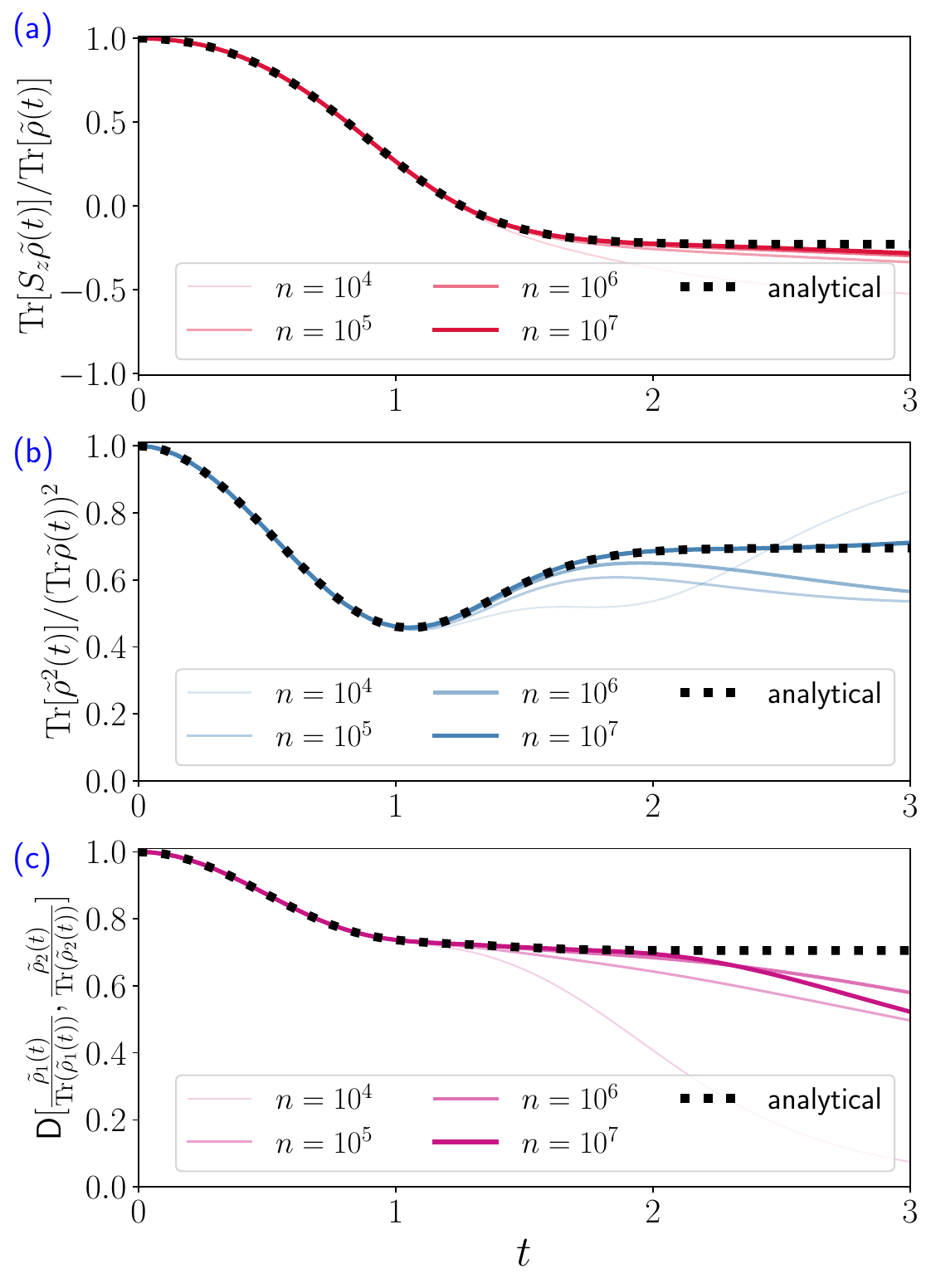}
\caption{(a) The magnetization of the disorder-averaged qutrit state reaches a plateau for Gaussian disorder, whereas (b) the purity reaches a dip around $t \simeq 1$, followed by an increase and a final plateau. However, the dynamical map being non-unital, it does not witness non-Markovianity. (c) The trace distance does not witness non-Markovianity, as it decays monotonically, without any revival.
Beyond very early times, for all the quantities of magnetization, purity, and trace distance, the numerical average converges to the analytical solution only when using a large number of disorder realizations. }
    \label{fig:qutrit_purity}
\end{figure}

In Fig.~\ref{fig:qutrit_purity}a, we plot the expectation value of a Hermitian observable, the spin-1 magnetization $S_z={\rm Diag}(1,0,-1)$ [see Eq.~\eqref{eq: spin_1 hamiltonian}] with an initial statevector $\ket{\psi_0}=(1, 0, 0)^T$. 
Its time-evolution can be analytically expressed as
\begin{equation}\label{eq: qutrit_magnetization}
    \frac{{\rm Tr}( S_z \tilde{\rho}(t))}{{\rm Tr}(\tilde{\rho}(t))}=\frac{12+2G_1-2^{5/3}G_3}{15+e^{\frac{3}{2} \sigma^2 t^2}+2^{\frac{4}{3}}G_2+2^{\frac{2}{3}}G_3} \,,
\end{equation}
which can be compared to the evolution of $\sigma_z$ in the Hermitian qubit case, see Eq.~\eqref{eq: qubit_magnetization}. 
Again, the evolution averaged over the Gaussian distribution decreases monotonically. The numerical average over $10^4$ trajectories agrees qualitatively with the analytical prediction for the average, but even at $10^7$ independent samples slight deviations are still visible at the longest times of $t\sim3$ considered.

Moreover, we calculate the purity of the normalized state~\cite{sergi2013non}, 
\begin{equation}
\begin{aligned}
\frac{\text{Tr}[\tilde{\rho}^2(t)]}{(\text{Tr}\tilde{\rho}(t))^2} = 
& 3  \Big(81 -6G_1 +6\cdot 2^{1/3}G_2 
 +5G_1^2 +4\cdot 2^{2/3}G_2^2 \\ 
& +6 \cdot 2^{1/3} G_3^2 -2^{4/3}G_1 G_2 -4G_2G_3 -2^{8/3}G_1G_3 \Big)\\
& / \left(15+e^{\frac{3}{2} \sigma^2 t^2}+ 2^{\frac{4}{3}}  G_2+2^{\frac{2}{3}} G_3\right)^2
\,.
\label{eq:purity_qutrit}
\end{aligned}
\end{equation}
As we can see in Fig.~\ref{fig:qutrit_purity}b, the purity has a dip around $t \simeq 1$, after which it rises again to finally saturate in a plateau (we have checked that the dip does not derive from the normalization, which monotonically increases in time as an exponential). This behavior is unlike the qubit case, where the purity of the averaged state monotonically decays. In contrast to the qubit case, the dynamical map generated by the non-Hermitian qutrit Hamiltonian ensemble is non-unital (see Appendix~\ref{app: unitality}), meaning that the non-monotonic behavior of the purity in this case is not a good witness for non-Markovianity. To study a potential non-Markovianity, we also plot the trace distance in Fig.~\ref{fig:qutrit_purity}c. Its monotonous decay with time is compatible with a Markovian evolution. 

These quantities do, however, illustrate an interesting aspect: Even at $10^6$ independent disorder samples, considerable deviations become visible already at around $\sim 1.75-2$. With $10^7$ samples, the purity reaches good convergence during the time frame considered, while the trace distance continues to display significant deviations.  
As this example shows, an extremely large number of realizations can be needed in order to match the numerical ensemble average to the analytical predictions, even for rather short times. 
Thus, even in simple systems it can be a considerable asset to be able to analytically extract converged results without the need for extensive numerical averaging. 

\section{Case III: $\hat{H}^3=h^2 \hat{H}$ }\label{sec: case III spin-1}
In the examples above, we have considered cases where finite powers of the Hamiltonian return to $\mathbb{I}$. 
As mentioned in Sec.~\ref{sec:finiteordergeneral}, we can derive similar analytical predictions also in the case when $\hat{H}^p=h^q \hat{H}^q$, with $q<p$. 
In this section, we illustrate this possibility for $\hat{H}^3=h^2 \hat{H}$, i.e., $p=3$ and $q=1$. 
The simplest instance of this case is given by $3\times 3$ matrices describing spin-1 operators, $S_{\alpha}$, $\alpha=x,y,z$. 
They follow the commutation relations $[S_{\alpha}, S_{\beta}]=i\hbar \epsilon_{\alpha \beta \gamma} S_\gamma$~\cite{mweene1999vectors}, and we choose the matrix representation
\begin{align}
    S_x=\frac{1}{\sqrt{2}} 
    \begin{pmatrix}
    0 & 1 & 0 \\
    1 & 0 & 1 \\
    0 & 1 & 0
    \end{pmatrix},
    S_y=\frac{1}{\sqrt{2}}
    \begin{pmatrix}
        0 & -i & 0 \\
        i & 0 & -i \\
        0 & i & 0
    \end{pmatrix},
    S_z=\begin{pmatrix}
        1 & 0 & 0 \\
        0 & 0 & 0 \\
        0 & 0 & -1 
    \end{pmatrix} \,.
\end{align}
Interestingly, these spin-$1$ operators also form a universal basis for qutrit computation~\cite{deller2023quantum, ekstrom2023variational,bottarelli2024inequality}. However, they follow a different symmetry than the qutrit or Potts operators discussed in Sec.~\ref{app:derivation_qutrit}.  
In particular, they are Hermitian, and they obey $(S_{\alpha})^3=S_{\alpha}$. 
It is, therefore, interesting to investigate the difference in the time-evolution these operators generate after disorder averaging. 

In similarity to the qubit case, Eq.~\eqref{eq: H_1},  we choose a Hamiltonian describing a rotation around a specific axis, 
\begin{equation}\label{eq: spin_1 hamiltonian}
   \hat{H}=\frac{h}{\sqrt{3}}(S_x+S_y+S_z) \Rightarrow \hat{H}^3=h^2\hat{H}\,.
\end{equation}
Using similar techniques to the previous cases, we can write down the final equation (see App.~\ref{app:derivation_spin_1}) 
\begin{equation}\label{eq: spin-1 final equation}
 \begin{aligned}
    \tilde{\vec{\rho}}(t)=& \Bigg[ \mathbb{I} \otimes \mathbb{I} + \left(G'-1\right) \Big\{ \tilde{H}^{*2} \otimes \mathbb{I} + \mathbb{I} \otimes  \tilde{H}^{2} \Big\} \\
    & + \frac{1}{2}\left(1-G\right)  \tilde{H}^* \otimes \tilde{H}+ \frac{1}{2}\left(3+G-4G' \right)\tilde{H}^{*2} \otimes \tilde{H}^2 \Bigg] \vec{\rho}(0)\,,
  \end{aligned}
\end{equation}
where  $G$ is identical to the qubit case (Tab.~\ref{tab:G(t)_qubit}), and  $G'$ has a similar form as $G$ but with $t\to t/2$, see Tab.~\ref{tab: all_propagator} for details.

\begin{figure}[hbt!]
    \centering
    \includegraphics[width=1\columnwidth]{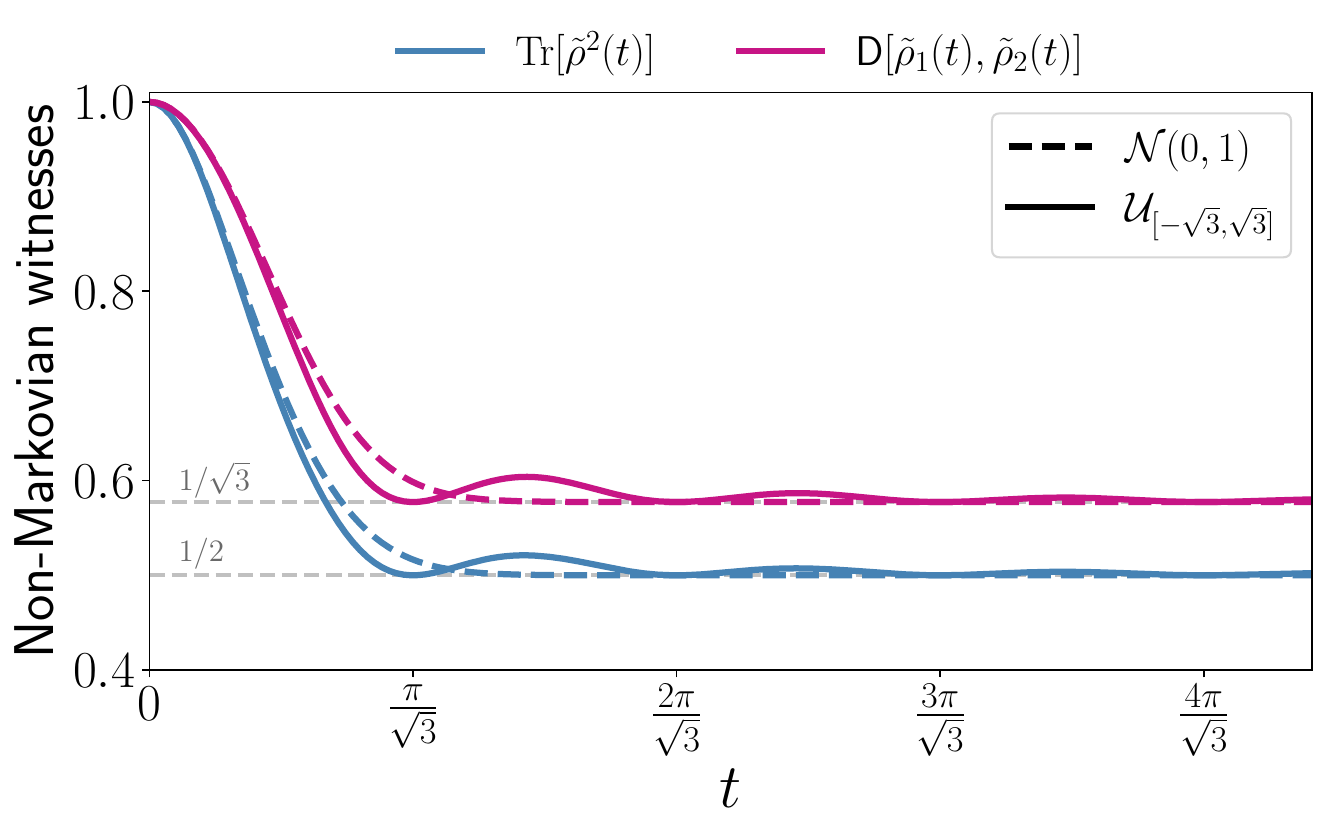}
    \caption{Time evolution of purity and trace distance of the disorder-averaged state for Gaussian (dashed line) and uniform (solid line) distribution. The time scales governing the evolution are twice those of the corresponding qubit dynamics (Fig.~\ref{fig:plot H1}).}
    \label{fig:plot for spin_1}
\end{figure}

Choosing the initial statevector $\ket{\psi_0}=(1, 0, 0)^T$, the expectation value of $S_z$ and the purity take the following forms: 
\begin{equation}
    \begin{aligned}
        {\rm Tr}(S_z\tilde{\rho} (t)) & = \frac{1}{3} \left(1+2G'(t)\right)\,,\\
        {\rm Tr}(\tilde{\rho}^2(t)) & = \frac{1}{18} \left(9+G^2(t) +8G'^2(t)\right)\,.
    \end{aligned}
\end{equation}
Figures~\ref{fig:intro_fig} and ~\ref{fig:plot for spin_1} respectively compare for both quantities the time evolution for the Gaussian and uniform distribution. 
For the Gaussian distribution, both observables monotoneously decay to a plateau. The one of the magnetization, ${\rm Tr}(S_z\tilde{\rho} (t\to\infty))=1/3$,  matches with the long-time plateau obtained in the qubit case, see Eq.~\eqref{eq: qubit_magnetization}, although the approach towards it is delayed, as it is governed by $G'(t)$ in the case of spin-$1$ instead of $G(t)$ as was the case for qubits. 
For the uniform distribution, both the magnetization and purity of the disorder-averaged state oscillate, with a period that is twice that of the qubit case (see Fig.~\ref{fig:intro_fig} for magnetization, and Fig.~\ref{fig:plot H1} 
and ~\ref{fig:plot for spin_1} for purity). In the long time limit, the oscillation dies out, and the dynamics approaches the plateau obtained in the case of Gaussian disorder. 
As discussed in Sec.~\ref{sec:nonmarkovianwitnesses}, as the map in the present case is unital (see Appendix \ref{app: unitality}), the oscillations of the purity indicate an emergent non-Markovianity of the averaged time evolution.

\section{Conclusion}\label{sec: conclusion}
In this work, we have derived an exact equation for the disorder-averaged dynamics of periodic Hamiltonians, independent of the initial states, applicable for Hermitian and non-Hermitian systems, and valid for arbitrary evolution times. These equations rely not on the specific representation of the Hamiltonians but rather only on their $(q,p)$-potency class as defined by the relation $\hat{H}^p \propto \hat{H}^q$. The effect of the disorder is then entirely captured in a few analytic functions derived from the disorder moments, which can be easily exchanged in the final equations according to the disorder at hand. 
This analytic approach significantly reduces the numerical overhead typically associated with disorder averaging. As often occurs in such a scenario~\cite{gneiting2016incoherent,kropf2016effective,chen2022effects,lu2023sudden,gneiting2020disorder,kropf2020protecting,chen2018simulating,han2019helical,araki2023robust}, the disorder averaging induces open-system-like dynamics even in inherently closed systems. In cases where the equation defining the time-evolved state can be inverted, we have also derived a master equation governing the disorder-averaged system dynamics.  

Moreover, even without a formal master equation, we have shown how non-Markovian characteristics emerge as a function of the disorder distribution and its strength. These factors determine the periodicity of revivals observed in non-Markovian witnesses. Such phenomena could be leveraged when using an ensemble of disordered Hamiltonians to simulate open system dynamics~\cite{chen2018simulating, mahdian2020incoherent} as well as for applications in quantum-information processing~\cite{
xu2013experimental,deffner2013quantum,bylicka2014non,reich2015exploiting,gneiting2020disorder,araki2023robust}. 
It is an interesting question to ask what it means to witness non-Markovianity in a closed system without a genuine bath, and how the associated backflow of information should be interpreted.

The analytic expressions may help forming a better understanding of errors in quantum computing by enabling insights into noise modeling or aiding in reverse-engineering disorder distributions tailored for specific applications. 
Another application can be in qudit-based quantum computations, where the effect of the disorder depends not only on the hardware implementation but also on the specific representation of the qudit operator. It thus becomes crucial to investigate which combinations of qudit dimension $d$, related to $q,p$, and disorder distributions $P(h)$ offer better resilience against decay caused by averaging over uncontrolled disorder in the qudit gates.  
Finally, in the context of optimal control, the optimal gate pulses for noisy hardware can be modified, knowing the decay rate caused by specific errors~\cite{de2024pulse}. 
Beyond an intrinsic interest in analytic solutions to disorder problems, such investigations could thus guide the development of more robust quantum computing systems.

\section{Acknowledgments}
We thank Alberto Biella, Soumik Bandyopadhyay, Alessio Pavliglianti, and Julius Mildenberger for useful discussions. This project has received funding by the European Union under Horizon Europe Programme - Grant Agreement 101080086 - NeQST and under NextGenerationEU via the ICSC – Centro Nazionale di Ricerca in HPC, Big Data and Quantum Computing. This project has received funding from the Italian Ministry of University and Research (MUR) through the FARE grant for the project DAVNE (Grant R20PEX7Y3A), was supported by the Provincia Autonoma di Trento, and by Q@TN, the joint lab between the University of Trento, FBK-Fondazione Bruno Kessler, INFN-National Institute for Nuclear Physics and CNR-National Research Council. 
Views and opinions expressed are however those of the author(s) only and do not necessarily reflect those of the European Union or the European Commission. Neither the European Union nor the granting authority can be held responsible for them. 

\appendix
\section{Derivation of the dynamical equation} \label{app: first equation derivation}
By series expanding the evolved density operator $\rho(t)$, Eq.~\eqref{eq:rhot_mat}, and vectorizing the operators according to Eq.~\eqref{eq:superoperator}, we reach the vectorized time-dependent equation for $\vec{\rho}(t)$, Eq.~\eqref{eq:rhot_vec}.
Specifically, 
\begin{align*}
  \rho(t) & = e^{-iHt} \rho(0)  e^{+i H^\dagger t} \\
  &= \sum_{k=0}^\infty \frac{(-it)^k}{k!} H^k \rho(0)  \sum_{j=0}^\infty \frac{(it)^j}{j!} (H^\dagger)^j  \\
   &= \sum_{n=0}^\infty \frac{(-it)^n}{n!} \sum_{k=0}^n \frac{n!}{(n-k)! k!} H^{k} \rho(0) (-H^\dagger)^{n-k} \,.
\end{align*}
Now, we can go to the super-operator notation by using Eq.~\eqref{eq:superoperator} to take the initial state towards the right and extract an evolution term that is independent of the initial state. Note that, the term $((-H^\dagger)^{n-k})^T=((-H^\dagger)^T)^{n-k}=(-H^*)^{n-k}$, and thus we obtain 
\begin{align} 
   \vec{\rho}(t)&= \sum_{n=0}^\infty \frac{(-it)^n}{n!} \sum_{k=0}^n {n \choose k} (\mathbb{I} \otimes H^{k}) ((-H^*)^{n-k} \otimes \mathbb{I}) \vec{\rho}(0) \nonumber\\
   &= \sum_{n=0}^\infty \frac{(-it)^n}{n!} \sum_{k=0}^\infty {n \choose k} (-H^*)^{n-k} \otimes H^k  \vec{\rho}(0) \,,  \label{eq:expanded}
\end{align}
which is nothing else than Eq.~\eqref{eq:rhot_vec}.  

\section{Derivations for Case I: $H^2 \propto \mathbb{I}$} 
\label{app:derivation_qubit}
In this section, we show how, starting from the general equation Eq.~\eqref{eq:rhot_vec} and using the symmetry of the Hamiltionian $H^2 \propto \mathbb{I}$, we reach the time-evolved state given in Eq.~\eqref{eq:rho_qubit} and the master equation~\eqref{eq: qubit master equation}. 

\paragraph{General time-evolved state:} 
To start with, we note that only terms even in $t$ (i.e., even in $h$) in Eq.~\eqref{eq:expanded} will survive disorder averaging. So, we only need to take care of the even $n$ terms. Splitting these into even and odd $k$, we need to analyze
\begin{equation*}
    \begin{aligned}
        &\bigg[ \sum_{k=0}^{2n} {2n \choose k} (-H^*)^{2n-k} \otimes H^{k} \vec{\rho}(0) \bigg] \\
        &= \bigg[ \sum_{k=0}^{n}  {2n \choose 2k} (-H^*)^{2n-2k} \otimes H^{2k} \\
        & \quad \quad \quad \quad + \sum_{k=0}^{n-1}  {2n \choose 2k+1} (-H^*)^{2n-2k-1} \otimes H^{2k+1}   \bigg]\ \vec{\rho}(0)\\
        &=\bigg[ h^{2n}  \mathbb{I} \otimes \mathbb{I}  \Big( \frac{2^{2n}+{\delta_{n0}}}{2} \Big)  - h^{2n-2} \tilde{H}^* \otimes \tilde{H} \Big(\frac{2^{2n}}{2} \Big)   \bigg] \ \vec{\rho}(0)\,.
    \end{aligned}
\end{equation*}
Note that the entire simplification of the terms is due to the periodicity of the Hamiltonian. Now, we can perform the average over the disorder, 
\begin{equation*}
    \begin{aligned}
         \tilde{\vec{\rho}}(t) = & \bigg[\sum_{n=0}^\infty  \frac{(-it)^{2n}}{(2n)!} \mathbb{E}[h^{2n}]  \Big( \frac{2^{2n}+{\delta_{n0}}}{2} \Big)  \mathbb{I} \otimes \mathbb{I} \\ & -   \sum_{n=1}^\infty  \frac{(-it)^{2n}}{(2n)!} \mathbb{E}[h^{2n}] \Big(\frac{2^{2n}}{2} \Big) \tilde{H}^* \otimes \tilde{H} \bigg] \ \vec{\rho}(0) \,.
    \end{aligned}
\end{equation*}
From here, using the moments of the disorder distributions, we will derive the exact analytical form of the equation Eq.~\eqref{eq:rho_qubit}

\paragraph{Gaussian disorder:}
In the special case of a Gaussian distribution $(0,\sigma^2)$, we have $\mathbb{E}[h^{2n}]=\sigma^{2n}(2n-1)!!$, and therefore 
\begin{equation}\label{eq: eq_app_gaussian moments}
    \frac{(-it)^{2n}}{(2n)!} \mathbb{E}[h^{2n}]
    = \frac{(-i\sigma t)^{2n}}{(2n)!!}= \frac{1}{n!}  \left( \frac{-\sigma^2 t^2}{2} \right)^n \,.
\end{equation}
Thus, we obtain
\begin{equation*}
    \sum_{n=0}^\infty \frac{1}{n!}  \left( \frac{-\sigma^2 t^2}{2} \right)^n \Big( \frac{2^{2n}+{\delta_{n0}}}{2} \Big)= \frac{1}{2} \Bigg[ 1 + e^{-2\sigma^2 t^2}   \Bigg]\,
\end{equation*}
and 
\begin{equation*}
     \sum_{n=1}^\infty \frac{1}{n!}  \left( \frac{-\sigma^2 t^2}{2} \right)^n \Big( \frac{2^{2n}}{2} \Big) = \frac{1}{2} \Bigg[ e^{-2\sigma^2 t^2} -1 \Bigg] \,.
\end{equation*}
The final equation for the average time-evolved state in the case of Gaussian disorder thus becomes
\begin{equation*}
    \tilde{\vec{\rho}}(t)=\frac{1}{2}\Bigg[ \left(1+e^{-2\sigma^2t^2}\right) \mathbb{I} \otimes \mathbb{I}+\left(1- e^{-2\sigma^2t^2}\right) \tilde{H}^* \otimes \tilde{H}  \Bigg]\ \vec{\rho}(0)\,.
\end{equation*}

\paragraph{Uniform disorder:}
For a uniform box distribution in the interval $[-b,b]$, we have $\mathbb{E}[h^{2n}]=b^{2n}/(2n+1)$. Thus, 
\begin{equation}\label{eq: eq_app_uniform moments}
    \frac{(-it)^{2n}}{(2n)!} \mathbb{E}[h^{2n}] =  \frac{(-ibt)^{2n}}{(2n+1)!} \,.
\end{equation}
Hence, we obtain 
\begin{equation*}
    \sum_{n=0}^\infty \frac{(-ibt)^{2n}}{(2n+1)!}\Big( \frac{2^{2n}+{\delta_{n0}}}{2} \Big) = \frac{1}{2} \Bigg[ 1+ \frac{\sin{2bt}}{2bt}  \Bigg]\,.
\end{equation*}
In this case, the disorder-averaged time-evolved state is
\begin{equation*}
    \tilde{\vec{\rho}}(t)=\frac{1}{2}\Bigg[ \left(1+ \frac{\sin{2bt}}{2bt} \right) \mathbb{I} \otimes \mathbb{I}+\left(1- \frac{\sin{2bt}}{2bt}\right) \tilde{H}^* \otimes \tilde{H}  \Bigg]\ \vec{\rho}(0)\,.
\end{equation*}
Again, $G(t)$ is given in Tab.~\ref{tab:G(t)_qubit}.

\paragraph{Deriving the master equation:}
From the analytic expression of the time-evolved state, given above and in Eq.~\eqref{eq:rho_qubit}, we can also obtain a von-Neumann type evolution equation for the disorder-averaged $\tilde{\rho}(t)$. We rewrite Eq.~\eqref{eq:rho_qubit} in terms of $\hat{A}$ and $\hat{B}$ as defined below,

\begin{equation}
\begin{aligned}
    \tilde{\vec{\rho}}(t) & = \frac{1}{2}\left[ \underbrace{\left( \mathbb{I}\otimes \mathbb{I} +\tilde{H}^*\otimes\tilde{H} \right)}_{\hat{A}}+ G(t) \underbrace{\left(\mathbb{I}\otimes \mathbb{I} -\tilde{H}^*\otimes\tilde{H}  \right)}_{\hat{B}}\right]\vec{\rho}(0) \\
    \Rightarrow \Lambda_t& =\frac{1}{2} \left(\hat{A}+G(t)\hat{B}\right)
\end{aligned}
\end{equation}

Following the steps in Eq.~\eqref{eq: lindbladian}, we start by taking the time derivative of $\tilde{\vec{\rho}}(t)$, 
\begin{equation}\label{eq:A}
    \partial_t \tilde{\vec{\rho}}(t)=\frac{\partial_t G(t)}{2} \hat{B} \vec{\rho}(0)\,.
\end{equation}
To replace $\vec{\rho}(0)=\Lambda_t^{-1} \vec{\tilde{\rho}}(t)$, we evaluate the inverse, which has the form,

\begin{equation}
    \Lambda_t^{-1}=\frac{1}{2}\left(\hat{A}+ \frac{\hat{B}}{G(t)} \right)\,,
\end{equation}
which follows from the fact that, $\tilde{H}^2=\mathbb{I}$ and thus $\hat{A}^2=2\hat{A},\hat{B}^2=2\hat{B}, \hat{A}\hat{B}=\hat{B}\hat{A}=0$, as well as from $\Lambda_0=\mathbb{I}$ whence $\frac{1}{2}(\hat{A}+\hat{B})=\mathbb{I}$. Thus, finally we obtain,

\begin{equation}
    \partial_t \tilde{\vec{\rho}}(t)= \frac{\partial_t G(t)}{2G(t)} \hat{B} \tilde{\vec{\rho}}(t)\,.
\end{equation}
By replacing $\hat{B}=(\mathbb{I}\otimes \mathbb{I}-\tilde{H}^* \otimes \tilde{H})$ and going back from super-operator to matrix notation, we obtain Eq.~\eqref{eq: qubit master equation}.

\section{Derivation for Case II: $\hat{H}^3 \propto \mathbb{I}$} 
\label{app:derivation_qutrit}
For the case $\hat{H}^3 = h^3 \mathbb{I}$, we can write more generally $H^k= h^{k} \tilde{H}^{k \pmod{3}}$. As before, the terms in Eq.~\eqref{eq:final eq} that are odd in $h$ vanish, and thus we are left with
\begin{equation*}
    \begin{aligned}
        &\bigg[ \sum_{k=0}^{2n} {2n \choose k} (-H^*)^{2n-k} \otimes H^{k} \vec{\rho}(0) \bigg] \\
        &= \bigg[ \sum_{k=0}^{2n} {2n \choose k} h^{2n}(-1)^{2n-k} \tilde{H}^{*(2n-k)\pmod{3}} \otimes \tilde{H}^{k\pmod{3}} \vec{\rho}(0) \bigg]\,.
    \end{aligned}
\end{equation*}
As $\tilde{H}^3 \propto \mathbb{I}$, the dynamics is generated by $\{\mathbb{I}, \tilde{H}, \tilde {H}^2\}$ and its complex conjugates, thus when written in super-operator formalism like $B^* \otimes A,$ where $A,B \in \{\mathbb{I},\tilde{H}, \tilde{H}^2\}$, there are $9$ possible terms. 

Now, we can use $\omega^3=1,$ to have a closed form of the combinatorial terms, 
\begin{widetext}
\noindent
\begin{equation}\label{eq: qutrit_dev_sum_combinatorial}
\begin{aligned}
     {n \choose 0} - {n \choose 3}+ {n \choose 6}......  &=\frac{0^n + (1-\omega)^n +(1-\omega^2)^n}{3}  
 \equiv W(n,1) \,,\\
     -{n \choose 1} + {n \choose 4}- {n \choose 7}......  &=\frac{1}{3}\left(\frac{\delta_{n 0}}{1}+\frac{(1-\omega)^n}{\omega}+\frac{(1-\omega^2)^n}{\omega^2}\right)\equiv W(n,\omega ) \,,\\
     {n \choose 2} - {n \choose 5}+ {n \choose 8}......  &=\frac{1}{3}\left(\frac{\delta_{n 0}}{1}+\frac{(1-\omega)^n}{\omega^2}+\frac{(1-\omega^2)^n}{\omega}\right)\equiv W(n,\omega^2 ) \,.
\end{aligned}
\end{equation}                                     
Using the above simplifications, the disorder-averaged density matrix is 
\begin{equation} 
\begin{aligned}
     \tilde{\rho}(t) = \sum_{n=0}^\infty &\Bigg[   \frac{(-it)^{6n}}{(6n)!} \mathbb{E}[h^{6n}] \Big\{ W(6n,1)\mathbb{I}\otimes \mathbb{I} + W(6n,\omega) {\tilde{H}^{*2}\otimes \tilde{H}} + W(6n,\omega^2) {\tilde{H}^* \otimes \tilde{H}^2}\Big\} \\
     & +\frac{(-it)^{6n+2}}{(6n+2)!}\mathbb{E}[h^{6n+2}] \Big\{ W(6n+2,1) \tilde{H}^{*2}\otimes \mathbb{I} + W(6n+2,\omega) \tilde{H}^{*}\otimes \tilde{H} + W(6n+2,\omega^2) \mathbb{I} \otimes \tilde{H}^2\Big\}\\
     &+ \frac{(-it)^{6n+4}}{(6n+4)!} \mathbb{E}[h^{6n+4}] \Big\{W(6n+4,1) {\tilde{H}^{*}\otimes \mathbb{I}} + W(6n+4,\omega) {\mathbb{I}\otimes \tilde{H}} +W(6n+4,\omega^2) \tilde{H}^{*2} \otimes \tilde{H}^2 \Big\}\Bigg]\ \vec{\rho}(0) \,.
\end{aligned}
\end{equation}
Using $(1-\omega)^{6n} =(-3)^{3n}$ in Eq.~\eqref{eq: qutrit_dev_sum_combinatorial}, we can further simplify the $W(6n, \cdot )$ terms,  
\begin{align}\label{eq:W_values}
    & W(6n,1)=\frac{\delta_{n0}+2(-3)^{3n}}{3}, \quad W(6n,\omega)=W(6n,\omega^2)=\frac{\delta_{n0}-(-3)^{3n}}{3}, \nonumber \\
    &W(6n+2,1)=W(6n+2,\omega^2)=\frac{-(-3)^{3n+1}}{3}, \quad 
     W(6n+2,\omega)=\frac{2(-3)^{3n+1}}{3}, \nonumber \\
    & W(6n+4,1)=W(6n+4,\omega)= \frac{-(-3)^{3n+2}}{3}, \quad 
    W(6n+4,\omega^2)=\frac{2(-3)^{3n+2}}{3} \,.
\end{align}
\end{widetext}
Note, the $\delta_{n0}$ terms in the coefficients corresponding to the operators $\mathbb{I} \otimes \mathbb{I}, \tilde{H}^{*2} \otimes \tilde{H}, \tilde{H}^* \otimes \tilde{H}^2$ are responsible for the time-independent terms in Eq.~\eqref{eq:qutrit_rhot}. The factor of $2$ in front of $\mathbb{I} \otimes \mathbb{I}$, $\tilde{H}^* \otimes \tilde{H}$ and $\tilde{H}^{*2} \otimes \tilde{H}^2$, and $-1$ in all the other terms of the time-dependent part are also due to the above relation in Eq.~\eqref{eq:W_values}.

Now, we need to simplify the time-dependent term along with disorder-averaging $\mathbb{E}[h^{\#}]$. In the case of uniform disorder, the result can be expressed in terms of hypergeometric function. For Gaussian disorder, using $\mathbb{E}[h^{2n}]=\sigma^{2n}(2n-1)!!$, one obtains rather compact expressions such as 
\begin{align*}
    &\sum_{n=0}^\infty  \frac{(-it)^{6n}}{(6n)!} \mathbb{E}[h^{6n}] =\sum_{n=0}^\infty \frac{(-it)^{6n}}{(6n)!}\sigma^{6n}(6n-1)!! \nonumber \\ &=\sum_{n=0}^\infty \frac{(-i\sigma t)^{6n}}{(6n)!!}  
    = \sum_{n=0}^\infty \frac{(-\sigma^2t^2)^{3n}}{2^{3n}(3n)!}= \sum_{n=0}^\infty \frac{(-\frac{1}{2}\sigma^2t^2)^{3n}}{(3n)!}\,               
\end{align*}
and similarly for the terms with $3n$ replaced by $3n+1$ and $3n+2$. The resulting infinite series can be expressed as the sum of three exponentials having permutations of $(1,\omega, \omega^2)$ in the exponents, 

\begin{equation*}
  \begin{aligned}
     &  \sum_{n=0}^\infty \frac{x^{3n+j}}{(3n+j)!}=\frac{e^x +\omega^{2j}e^{\omega x}+\omega^{4j}e^{\omega^2 x}}{3} \,,  \quad \text{for} \
      j=0,1,2 \,,
  \end{aligned}
\end{equation*}
These relations allow us to express the three time-dependent functions $G_1, G_2, G_3$ in terms of $\vec{v}=(e^x, e^{\omega x}, e^{\omega^2 x})$ and the three vectors $(1,1,1)^T$, $(1,\omega^2, \omega)^T$, $(1,\omega, \omega^2)^T$, 
leading to the Eq.~\eqref{eq:qutrit_rhot} describing the disorder-averaged time evolution of a qutrit.

\section{Derivation for Case III: $\hat{H}^3 \propto \hat{H}$} 
\label{app:derivation_spin_1}

As in the previous cases, terms of odd order in $h$ vanish. Thus, starting from the Eq.~\eqref{eq:final eq}, we are left with 
\begin{equation*}
    \begin{aligned}
        \sum_{n=0}^\infty \frac{(-iht)^{2n}}{(2n)!}  \bigg[ \sum_{k=0}^{2n} {2n \choose k} (-\tilde{H}^*)^{2n-k} \otimes \tilde{H}^{k} \bigg]\,.
    \end{aligned}
\end{equation*}
One can consider the terms $n=0,1,2, \dots$ one by one and, by putting the terms together, can find the following equation:
 
\begin{widetext}
\begin{equation*}
   \mathbb{I} \otimes  \mathbb{I} + \sum_{n=1}^{\infty} \frac{(-iht)^{2n}}{(2n)!} \underbrace{{2n \choose 0}}_{=1} \tilde{H}^{*2} \otimes \mathbb{I} - \sum_{n=1}^{\infty} \frac{(-iht)^{2n}}{(2n)!} \underbrace{\sum_{k=0}^{n-1} {2n \choose 2k+1}}_{=2^{2n-1}}\tilde{H}^* \otimes \tilde{H} + \sum_{n=2}^\infty \frac{(-iht)^{2n}}{(2n)!} \underbrace{\sum_{k=1}^{n-1} {2n \choose 2k}}_{=2^{2n-1}-2}  \tilde{H}^{*2} \otimes \tilde{H}^2 + \sum_{n=1}^{\infty} \frac{(-iht)^{2n}}{(2n)!} \underbrace{{2n \choose 2n}}_{=1}  \mathbb{I} \otimes  \tilde{H}^{2}  
\end{equation*}   

For the Gaussian disorder case, using Eq.~\eqref{eq: eq_app_gaussian moments}, disorder averaging yields 
\begin{equation*}
     \mathbb{I} \otimes  \mathbb{I} + \left( {e^{-\sigma^2 t^2/2}-1} \right) \tilde{H}^{*2} \otimes \mathbb{I} - {\frac{1}{2} \left(e^{-2\sigma^2 t^2}-1\right)} \tilde{H}^* \otimes \tilde{H} + 
    \frac{1}{2}\left(3+e^{-2 \sigma^2 t^2} -4 e^{-\sigma^2 t^2/2} \right)
    \tilde{H}^{*2} \otimes \tilde{H}^2 + \left({e^{-\sigma^2 t^2/2}-1}\right) \mathbb{I} \otimes  \tilde{H}^{2} \,.
\end{equation*}
For the uniform disorder case, using Eq.~\eqref{eq: eq_app_uniform moments}, we obtain  
\begin{equation}
    \mathbb{I} \otimes  \mathbb{I} + \left({\frac{\sin{bt}}{bt}-1}\right) \tilde{H}^{*2} \otimes \mathbb{I} - \frac{1}{2}\left(\frac{\sin{2bt}}{2bt}-1\right)  \tilde{H}^* \otimes \tilde{H} + \frac{1}{2}\left(3+\frac{\sin{2bt}}{2bt}-4\frac{\sin{bt}}{bt} \right)\tilde{H}^{*2} \otimes \tilde{H}^2 + \left({\frac{\sin{bt}}{bt}-1}\right)  \mathbb{I} \otimes  \tilde{H}^{2}\,.
\end{equation}
\end{widetext}

Thus, we obtain the Eq.~\eqref{eq: spin-1 final equation} with two time-dependent functions $G(t)$ and $G'(t)$ as described in Tab.~\ref{tab: all_propagator}.

\section{Summary of the result}
In this appendix, we summarise the result for the three cases, namely qubits, qutrits with $\mathbb{Z}_3$, and spin-1 qutrits in the table below. Table~\ref{tab: all_propagator} contains the dynamical map $\Lambda_t$ corresponding to the disorder-averaged dynamics and the time-dependent decoherence function therein both for Gaussian and uniform disorder. We further summarize the average magnetization and non-Markovian witnesses in Table.~\ref{tab: all_observables} and ~\ref{tab: all_non_markovian}, respectively.

\begin{table*}
    \centering
    \renewcommand{\arraystretch}{1.7} 
    \begin{tabular}{|c||c|c|c|}
     \hline
     Hamiltonian ($\hat{H}=h \tilde{H}$) & 
     $\Lambda_t$ characterizing dynamics & Time-dependence for $\mathcal{N}(0,\sigma^2)$ 
     & Generators for $\mathcal{U}_{[-b,b]}$  
     \\ \hline \hline
     
     $\tilde{H}^2=\mathbb{I}$; $\tilde{H}^\dagger =\tilde{H}$ & 
     $\frac{1}{2}\big(\mathbb{I} \otimes \mathbb{I}+ \hat{H}^* \otimes  \hat{H}\big) + \frac{G(t)}{2} \big(\mathbb{I} \otimes \mathbb{I}- \hat{H}^* \otimes  \hat{H}\big)$ 
     & $G(t)=e^{-2 \sigma^2 t^2}$   
     & $G(t)= \frac{\sin{2bt}}{2bt}$\\ 
     \hline 

     $\tilde{H}^3= \mathbb{I}; \tilde{H}^\dagger \neq \tilde{H}$ & 
     \begin{tabular}[c]{@{}l@{}}
     $\frac{1}{3}\big(\mathbb{I} \otimes \mathbb{I}+{\tilde{H}^{*2}\otimes \tilde{H}} +  {\tilde{H}^{*}\otimes \tilde{H}^2} \big)$ \\  
     $+ \frac{G_1(t)}{9} \big(2\mathbb{I} \otimes \mathbb{I} -  \tilde{H}^{*2}\otimes \tilde{H} -\tilde{H}^* \otimes \tilde{H}^{2}\big)$ \\ 
     $+ \frac{G_2(t)}{9} \big(2{\tilde{H}^{*}\otimes \tilde{H}} -  \tilde{H}^{*2}\otimes \mathbb{I}- \mathbb{I} \otimes \tilde{H}^{2}\big)$ \\ 
     $+ \frac{G_3(t)}{9} \big(2{\tilde{H}^{*2}\otimes \tilde{H}^2} -  \tilde{H}^{*}\otimes \mathbb{I}   -\mathbb{I} \otimes \tilde{H}\big)$
     \end{tabular} & 
     \begin{tabular}[c]{@{}l@{}}
     $G_1(t)=e^{\frac{3}{2}\sigma^2 t^2}+e^{\frac{3}{2}\omega \sigma^2 t^2}+e^{\frac{3}{2}\omega^2 \sigma^2 t^2}$ \\ 
     $G_2(t)=e^{\frac{3}{2}\sigma^2 t^2}+\omega^2e^{\frac{3}{2}\omega \sigma^2 t^2}+\omega e^{\frac{3}{2}\omega^2 \sigma^2 t^2}$ \\ 
     $G_3(t)=e^{\frac{3}{2}\sigma^2 t^2}+\omega e^{\frac{3}{2}\omega \sigma^2 t^2}+\omega^2 e^{\frac{3}{2}\omega^2 \sigma^2 t^2}$
     \end{tabular} & --- \\ \hline 

     $\tilde{H}^3=\tilde{H}; \tilde{H}^\dagger =\tilde{H}$ &
     \begin{tabular}[c]{@{}l@{}}
     $(\mathbb{I} \otimes \mathbb{I}-\tilde{H}^{*2} \otimes \mathbb{I}-\mathbb{I} \otimes \tilde{H}^2)$ \\ 
     $+ \frac{1}{2}(\tilde{H}^* \otimes \tilde{H}+3\tilde{H}^{*2} \otimes \tilde{H}^2)$ \\ 
     $+ \frac{G(t)}{2}(\tilde{H}^{*2} \otimes \tilde{H}^2- \tilde{H}^* \otimes \tilde{H})$ \\ 
     $+ G'(t)(\tilde{H}^{*2} \otimes \mathbb{I}+\mathbb{I} \otimes \tilde{H}^2-2\tilde{H}^{*2} \otimes \tilde{H}^2)$
     \end{tabular} & 

     \begin{tabular}[c]{@{}l@{}}
     $G(t) = e^{-2\sigma^2 t^2}$ \\ 
     $G'(t) = e^{-\sigma^2 t^2 / 2}$
     \end{tabular} &  
     \begin{tabular}[c]{@{}l@{}}
     $G(t)=\frac{\sin{2bt}}{2bt}$ \\ 
     $G'(t)=\frac{\sin{bt}}{bt}$
     \end{tabular}
        \\ \hline 
    \end{tabular}
    \caption{The disorder-averaged dynamics is defined through Eq.~\eqref{eq:final eq} by the map $\Lambda_t$ (2nd column), which depends on different powers of $\tilde{H}$, and time-dependent functions that contain all the information about the disorder distribution. The exact forms of the time-dependent functions are given in the 3rd and last columns for Gaussian and uniform disorder, respectively. We omitted the unwieldy formula involving hypergeometric functions for the case $\tilde{H}^3= \mathbb{I}; \tilde{H}^\dagger \neq \tilde{H}$ and uniform disorder.}
    \label{tab: all_propagator}
\end{table*}

\begin{table*}
    \centering
    \renewcommand{\arraystretch}{1.7} 
    \begin{tabular}{|C{2.cm}||C{1.9cm}|C{3.8cm}|}
     \hline
     Hamiltonian & Initial state  & 
     Average magnetization 
     \\ \hline \hline
     
     $\frac{h}{\sqrt{3}}(\sigma_x+\sigma_y+\sigma_z)$ 
     & 
        $\ket{\uparrow}$  
     & 
     ${\rm Tr}[\sigma_z \tilde{\rho}(t)]=\frac{1}{3}(1+2G(t))$ 
    \\

     \hline 
  
     $\frac{h}{2^{1/3}}(\sigma+\tau)$ & 
     $\ket{S_z,+1}$ & 
     $ \frac{{\rm Tr}( S_z \tilde{\rho}(t))}{{\rm Tr}(\tilde{\rho}(t))}=\frac{12+2G_1-2^{5/3}G_3}{15+e^{\frac{3}{2} \sigma^2 t^2}+2^{\frac{4}{3}}G_2+2^{\frac{2}{3}}G_3} $
      \\ \hline 

     $\frac{h}{\sqrt{3}}(S_x+S_y+S_z)$ & 
      $\ket{S_z,+1}$ 
      & ${\rm Tr}(S_z\tilde{\rho} (t)) = \frac{1}{3} \left(1+2G'(t)\right)$ 
        \\ \hline 
    \end{tabular}
    \caption{Exact expressions for the evolution of disorder-averaged magnetization (column 3) starting from the initial state (column 2), expressed in terms of the decoherence functions (see Tab.~\ref{tab: all_propagator}).}
    \label{tab: all_observables}
\end{table*}

\begin{table*}[hbt!]
    \centering
    \renewcommand{\arraystretch}{1.7} 
    \begin{tabular}{|C{2.cm}||C{2.5cm}|C{3.35cm}|C{6cm}|}
     \hline
     Hamiltonian & Initial statevector  &  Non-Markovian witness
     & Exact expression of the witness
     \\ \hline \hline
     
     $\frac{h}{\sqrt{3}}(\sigma_x+\sigma_y+\sigma_z)$ & 
     \begin{tabular}[c]{@{}c@{}}
        $\ket{\uparrow}$  \\
        $\ket{\uparrow},\ket{\downarrow}$ \\
        $\frac{1}{\sqrt{2}}(\ket{\uparrow\uparrow}+\ket{\downarrow\downarrow})$
     \end{tabular}
      
     & 
     \begin{tabular}[c]{@{}c@{}}
        Purity: Tr$(\tilde{\rho}^2)$\\
        Trace distance: $D(\tilde{\rho}_1,\tilde{\rho}_2;t)$\\
        Log. negativity: $E_N(\tilde{\rho}_{SA})$
     \end{tabular}

     & 
     \begin{tabular}[c]{@{}c@{}}
      $\frac{1}{3}[2+G^2(t)]$    \\
      $\sqrt{\frac{1+2G^2(t)}{3}}$ \\
      $\log_2(1+|G(t)|)$
     \end{tabular}
    \\

     \hline 
  
     $\frac{h}{2^{1/3}}(\sigma+\tau)$ & 
     $\ket{S_z,+1}$
      
     & Purity: $\text{Tr}[\tilde{\rho}^2(t)]/(\text{Tr\ }\tilde{\rho}(t))^2$
      & $ 
        \frac{\begin{array}{c}
            3  \Big[81 -6G_1 +6\cdot 2^{1/3}G_2 +5G_1^2 +4\cdot 2^{2/3}G_2^2 \\ 
            +6 \cdot 2^{1/3} G_3^2 -2^{4/3}G_1 G_2 -4G_2G_3 -2^{8/3}G_1G_3 \Big]
                \end{array}
            }{
            \left(15+e^{\frac{3}{2} \sigma^2 t^2}+ 2^{\frac{4}{3}}  G_2+2^{\frac{2}{3}} G_3\right)^2}$
      
      \\ \hline 

     $\frac{h}{\sqrt{3}}(S_x+S_y+S_z)$ & 
     \begin{tabular}[c]{@{}c@{}}
        $\ket{S_z,+1}$   \\
        $\ket{S_z,+1},\ket{S_z,-1}$
     \end{tabular}

      & \begin{tabular}[c]{@{}c@{}}
        Purity: Tr$[\tilde{\rho}^2(t)]$   \\
        Trace distance: $D(\tilde{\rho}_1,\tilde{\rho}_2;t)$
        \end{tabular}
      & 
      \begin{tabular}[c]{@{}c@{}}
        $\frac{1}{18} \left(9+G^2(t) +8G'^2(t)\right)$ \\
        $\sqrt{\frac{1+2G'^2(t)}{3}}$
        \end{tabular}
      
        \\ \hline 
    \end{tabular}
    \caption{Exact expressions of non-Markovian witnesses (column 4) expressed in terms of the time-dependent parts $G(t)$ (see Tab.~\ref{tab: all_propagator}) arising in the dynamics of corresponding Hamiltonians (column 1).}
    \label{tab: all_non_markovian}
\end{table*}

\section{Unitality of the dynamical maps}\label{app: unitality}
A dynamical map is called unital when $\Lambda_t[\mathbb{I}]=\mathbb{I}$, i.e., a fully mixed state remains fully mixed. Here, $\Lambda_t$ acts on the density matrix, not on its vectorized form. 
We now show that the map for the qubit and spin-1 cases is unital, while it is non-unital for the non-Hermitian qutrit case considered. 

\paragraph{Case I:} For the qubit case, from Eq.~\eqref{eq:rho_qubit} we have
\begin{equation}
    \Lambda_t[\mathbb{I}]=\left( \frac{1+G(t)}{2}\right)\mathbb{I}+\left( \frac{1-G(t)}{2}\right) \tilde{H} \mathbb{I} \tilde{H}^\dagger=\mathbb{I}\,,
\end{equation}
where we used the fact that $\tilde{H}^\dagger=\tilde{H}$ and $\tilde{H}^2=\mathbb{I}$, showing that the dynamical map for the qubit system is unital.

\paragraph{Case II:} For the qutrit $(\mathbb{Z}_3)$ case, from Eq.~\eqref{eq:qutrit_rhot} we have
\begin{equation}
\begin{aligned}
    \Lambda[\mathbb{I}] & =\frac{1}{3} \big(\mathbb{I} + \tilde{H}{\tilde{H}}^{\dagger 2} +  \tilde{H}^2{\tilde{H}}^{\dagger}  \big)  
     +  \frac{G_1(t)}{9} \big(2\mathbb{I}  -  \tilde{H}\tilde{H}^{\dagger 2} - \tilde{H}^{2} \tilde{H}^{\dagger} \big) \\ 
     + & \frac{G_2(t)}{9} \big(2{\tilde{H}}\tilde{H}^{\dagger}  -  \tilde{H}^{\dagger 2}- \tilde{H}^{2}\big)  
     +  \frac{G_3(t)}{9} \big(2 {\tilde{H}^2} \tilde{H}^{\dagger 2}- \tilde{H}^{\dagger}  - \tilde{H}\big)\,.
\end{aligned}    
\end{equation}
As all the coefficients of $G_i(t)$ (in the bracket) remain non-zero, the dynamical map corresponding to qutrit is non-unital.

\paragraph{Case III:} For the spin-1 case, from Eq.~\eqref{eq: spin-1 final equation} we have 
\begin{equation}
  \begin{aligned}
       \Lambda_t[\mathbb{I}] & = \mathbb{I}+(G'-1)(\tilde{H}^{\dagger 2}+\tilde{H}^2) \\
       & +\frac{1}{2}(1-G)\tilde{H}\tilde{H}^\dagger+\frac{1}{2}(3+G-4G')\tilde{H}^2\tilde{H}^{\dagger 2}\,.
  \end{aligned}
\end{equation}

Note that $\tilde{H}^{\dagger 2}+\tilde{H}^2-2 \tilde{H}^2 \tilde{H}^{\dagger 2}=0$ and $\tilde{H}^2 \tilde{H}^{\dagger 2}-\tilde{H}\tilde{H}^\dagger=0$ as $\tilde{H}^\dagger=\tilde{H}$ and $\tilde{H}^3=\tilde{H}$. Thus, we can rewrite the above equation as 
\begin{equation}
    \Lambda_t[\mathbb{I}]=\mathbb{I}-(\tilde{H}^{\dagger 2}+\tilde{H}^2) + \frac{1}{2} \tilde{H}\tilde{H}^{\dagger}+\frac{3}{2}\tilde{H}^2\tilde{H}^{\dagger 2}=\mathbb{I}\,,
\end{equation}
and thus the dynamical map for the spin-1 system is unital.

\end{document}